# APPLICATIONS OF BIORTHOGONAL DECOMPOSITIONS IN FLUID-STRUCTURE INTERACTIONS

P. HÉMON

*Hydrodynamics Laboratory, LadHyX,*

*Ecole Polytechnique-CNRS, F-91128 Palaiseau Cedex, France*

*Email : pascal.hemon@ladhyx.polytechnique.fr*

and

F. SANTI

*CNAM, Department of Mathematics,*

*192, rue Saint-Martin, F-75141 Paris Cedex 03, France*

## ABSTRACT

This paper is dedicated to the study of the orthogonal decomposition of spatially and temporally distributed signals in fluid-structure interaction problems. First application is concerned with the analysis of wall-pressure distributions over bluff bodies. The need for such a tool is increasing due to the progress in data acquisition systems and in computational fluid dynamics. The classical proper orthogonal decomposition (POD) method is discussed, and it is shown that heterogeneity of the mean pressure over the structure induces difficulties in the physical interpretation. It is then proposed to use the biorthogonal decomposition (BOD) technique instead; although it appears similar to POD, it is more general and fundamentally different since this tool is deterministic rather than statistical. The BOD method is described and adapted to wall-pressure distribution, with emphasis on aerodynamic load decomposition. The second application is devoted to the generation of a spatially correlated wind velocity field which can be used for the temporal calculation of the aeroelastic behaviour of structures such as bridges. In this application, the space-time symmetry of the BOD method is absolutely necessary. Examples are provided in order to illustrate and show the satisfactory performance and the interest of the method. Extensions to other fluid-structure problems are suggested.

# 1. INTRODUCTION

In the last decade, the large increase in the power of computers and of multiple-channel measurement systems has led to a very large amount of collected data which have become difficult to analyse. In wind tunnel techniques, it is now common to encounter models equipped with hundreds of pressure taps. The same issues arise in computational aerodynamics. In this case, it becomes problematic to analyse physically the results, and it is necessary to process the data so as to extract physically meaningful information. The spatio-temporal complexity of the data gives rise to a variety of regimes, from random to periodic, and leads to very different probability density functions, from Gaussian to convex parabolic.

As pointed out in (Holmes *et al.* 1997), Armitt (1968) was probably the first in wind engineering to use an orthogonal decomposition for the wall-pressure distribution over a structure. This was performed by seeking the eigenmodes of a covariance matrix, but the actual term Proper Orthogonal Decomposition (POD) was not employed at that time. The purpose was not only to analyse the results, but also to create typical shapes associated with typical time histories for the structure under consideration.

It is interesting to note that more or less at the same time, in 1967, Lumley introduced the POD in order to extract coherent structures from the velocity field [see for instance (Holmes *et al.* 1996)]. These tools are now used in the field of dynamical systems and applied to aerodynamic flows (Cordier 1996). Another example is found in the paper by Delville *et al.* (1999), where measurements with rakes of hotwires are analysed. The authors complement their two component velocity field by using the continuity equation and the Taylor hypothesis in order to rebuild the missing third component. The analysis is carried out on the two point spectral tensor which is not different in spirit from the classical POD.

The problem of jet noise has been studied by Arndt *et al.* (1997). Since only a few microphones were available, a more complete acoustic pressure signal was obtained by using hypotheses such as stationarity and geometrical symmetry. The POD allowed the authors to measure the phase velocity, which is an important parameter in this application.

It must be recalled also that the method is not limited to fluid mechanical problems and it is also widely used in other communities. For instance, statisticians call the technique the principal component analysis (Lebart *et al.* 1982; Marriott 1974). In structural vibrations, Feeny & Kappagantu (1998) showed that the POD is equivalent to the modal expansion in terms of the classical linear eigenmodes of a structure.

However, although the POD method is very powerful for analysis of pressure distributions, published work remains rare and recent discussions have identified some problems in its implementation. The purpose of this paper is to clarify some points, related notably to the inclusion or not of the mean component (Tamura *et al.* 1999; Hémon & Santi 2002). When the mean pressure over the surface is heterogeneous, we shall indeed show that it introduces a bias in the POD analysis, independently of how the calculations are carried out, with or without inclusion of the mean value. It is also our intention to provide a more rigorous method of analysis, by using the Biorthogonal Decomposition (BOD) with application to fluid-structure interaction problems.

It will be proposed additionally to use the specific properties of the BOD technique in order to build a generator of a spatially correlated wind velocity field, which may be used in the temporal computation of the structural response to aeroelastic excitation. The application lies for instance in the nonlinear behaviour of large bridges under atmospheric turbulence excitation.

The paper is organized as follows. In the next section, we briefly present the POD method and we discuss its application and validity on bluff-body wall-pressure distribution. Section 3 is devoted to the presentation of the biorthogonal decomposition of Aubry *et al.* (1991) which we think is more appropriate for the problem at hand. Although it seems similar to the POD, this tool is deterministic instead of statistical. Some of the hypotheses regarding the analysed signal are no longer necessary. Adaptation to pressure distribution analysis is performed in this section. As pointed out by Dowell *et al.* (1999), the decomposition of the aerodynamic loads into eigenmodes contributes a “*friendly way*” for those who have to compute the structural response and even for the developers of active control systems. We apply the technique in Section 4 to two examples linked to the behaviour of bluff bodies. In Section 5, the BOD technique is used to develop a method for simulating a turbulent wind velocity field which can serve as an input for temporal aeroelastic computations.

## 2. THE PROPER ORTHOGONAL DECOMPOSITION

### *2.1. General presentation of the method*

The method is based on the Karhunen-Loève decomposition of a multivariable signal. The main idea is to find a set of proper orthogonal functions that capture the maximum of the energy of the signal with the minimum number of proper functions. The mathematical formulation can be found in many references, for instance in Holmes *et al.* (1996) and we present here only a summary. We assume that the data to analyse satisfy the required hypotheses, at least stationarity and ergodicity.

In this section the discrete wall pressure distribution $\boldsymbol{P}(\boldsymbol{x},t)$ around the structure is analysed. *N* pressure taps (*N* nodes) on the surface are to be analysed, each of them being simultaneously measured or computed for a sufficiently long time. *K* is the number of time steps. In what follows, $\langle r,s \rangle$ will be the Euclidian scalar product between $r$ and $s$, $\langle s \rangle$ the space integration of $s$ (over the surface of the structure for instance), and $\bar{s}$ denotes time integration of $s$.

#### *2.1.1. Direct method*

The direct POD method is formally written as

$$\boldsymbol{P}(\boldsymbol{x},t) = \sum_{n=1}^{N} \mu_n(t)\ \boldsymbol{W}_n(\boldsymbol{x}), \qquad (1)$$

where the proper functions are $\boldsymbol{W}_n$ and the principal components $\mu_n$. The optimisation problem of the Karhunen-Loève decomposition (1) consists in finding the best proper functions that maximize the mean square of the signal:

$$\max \left\{ \overline{\langle \boldsymbol{P},\boldsymbol{W} \rangle^2} \Big/ \langle \boldsymbol{W},\boldsymbol{W} \rangle \right\}. \qquad (2)$$

By using a Lagrangian formulation, it can be shown that this reduces to the eigenvalue problem

$$\boldsymbol{R}\,\boldsymbol{W} = \lambda\,\boldsymbol{W}\ , \qquad (3)$$

where $\boldsymbol{R}$ is the cross-correlation (or covariance) matrix of the pressure distribution. The use of the cross-correlation matrix including the mean value of the pressure, or the covariance matrix which is carried out on centred signals will be discussed in the next section.

The eigenvalues $\lambda_n$ are representative of the energy levels of each term. The proper functions benefit from the classical properties of the eigenmodes, mainly orthogonality and normalization.

Since the POD is optimal in energy, it becomes obvious that only the first terms in decomposition (1) will participate in the dynamics, so that only $M$ modes, with $M<<N$, have to be used. One of the great advantages of the POD is to reduce considerably the amount of data to be stored.

### *2.1.2. The snapshot method*

The snapshot method was suggested by Sirovich in 1987 [see for instance Breuer & Sirovich (1991) and references therein] in order to decrease the size of the eigenvalue problem (3). Indeed, when the data to analyse are obtained experimentally, the number of measurement points is generally small in comparison to the number of time steps, i.e. $N<<K$. The direct method will be preferred in this case, leading to an eigenvalue problem of dimension $N^2$.

However, in the analysis of computational results, the spatial resolution is generally very good and the simulation time is shorter, leading to $N>>K$. Then, by symmetry, Sirovich introduced the snapshot method in which the role of space is taken by time and conversely. The associated eigenvalue problem is only of dimension $K^2$. The proper functions now have a temporal significance, and the principal components represent the spatial information. The snapshot POD benefits from the same properties as the classical POD.

The direct and the snapshot methods were found to be very robust with respect to modifications in spatial or temporal resolution (Breuer & Sirovich 1991), which means that changes in the data-resolution have little effect on the accuracy of the proper functions. These authors also found that noise injected in the data does not perturb significantly the spectrum of the POD, since the eigenvalues are modified at a considerably lower level than the noise level.

## *2.2. Problems arising in bluff body wall-pressure analysis*

The use of the POD in wind engineering is now well accepted and some problems have been reported in different applications. In the paper by Holmes *et al.* (1997) the direct POD was used for problem involving a low-rise building with the wind normal to one of the vertical walls, starting from the correlation matrix including an internal pressure. The authors concluded that the constraints created by the orthogonality requirements were dominating the shape of the proper functions, so that no physical interpretation could be given to them. The only advantage of the POD was found to be an economical way of storing the data. The comparison between calculations including or not the vertical walls was performed and leads to these conclusions.

In another application on domes, Letchford & Sarkar (2000) utilized the POD method including mean components: they found that the second proper function was following the pressure gradient with respect to the wind direction, the mean component being given by the first proper mode. The main difference with the application on prismatic structures is the continuous progressive mean value evolution on domes, instead of steep changes on prismatic structures.

In our opinion, the problem arises in fact from the heterogeneity of the mean components, and not from their inclusion or exclusion in the analysis, as discussed by Tamura (1999) and Hémon & Santi (2002).

### *2.3.1. Geometrical interpretation*

For prismatic bluff bodies , the walls normal to the flow are subjected more or less (i) to a stagnation pressure value for the windward face, and (ii) to a complete stalled negative pressure for the downstream face. In 2 D state space, the clouds of the measured values are then located in different distinct regions, creating different clusters.

To illustrate this point, the pressure distribution over the rectangular section (Hémon & Santi 2002) is taken. This section has a length-to-thickness shape ratio of 2, and alternate vortex shedding is the dominant mechanism. A few nodes over the surface presented in Figure 1 have been selected as being good representatives of the signals. In Figure 2, we have plotted the clouds of the different state loci obtained for each nearest node, taken 2 by 2, i.e.

$$p_j = f(p_i) \quad \text{with} \quad j = i+1. \qquad (4)$$

One can see that the heterogeneous mean components fill the space in distinct parts, creating clusters around each different mean value encountered. The cluster around (1, 1) is linked to the front face of the rectangle, and the cluster around (-1, -1) to the lateral and rear sides.

In such a 2 D case, the POD method has usually a simple geometrical interpretation, since the equation

$$\left\langle \boldsymbol{P}^{\mathrm{t}}\ \boldsymbol{R}\ \boldsymbol{P} \right\rangle = c \qquad (5)$$

represents a set of ellipsoids centred on the origin. Then it is well known that seeking the principal components is equivalent to seeking the principal axes of the ellipsoid, in the order of length, because "*the calculations involved in finding the principal components are precisely those necessary to find the* *principal axes of the ellipsoid*" (Marriot 1974). This can easily be shown by developing the quadratic form (17) and searching for the ellipse axes.

Therefore, it is obvious that the proper functions obtained by analysing the data plotted in Figure 2 do not have a simple physical interpretation. Moreover, extracting the mean component leads to the same results because the four clusters will be centred on the origin and statistically mixed with each other, although their individual shape is very different. This problem can be reinforced by the nature of the individual pressure signals when a high level of periodicity is present.

*2.3.2. Probability density function analysis*

This point is in fact a major difference with some applications involving more or less random signals. In the case of the rectangular cylinder, the harmonic component is dominant, as shown in Figure 3 where typical samples have been selected. The probability density functions are really different from a Gaussian shape which makes the previous comments not completely applicable: the geometrical interpretation of the POD with ellipsoid and principal axes is indeed restricted to multivariate normal distributions with common mean, allowing only heterogeneity of variance (Marriot 1974). This is probably more or less the case in the application of Tamura (1999), but not at all for the present rectangular section.

Moreover, it is obvious (in 2 D) that harmonic signals, at the same frequency, generate an ellipse in the state plane, for which the rotation of the principal axes is directly a function of the phase angle between the two signals: when it is zero (modulus $\pi$), the signals are perfectly correlated and the ellipse reduces to a straight line. One of the eigenvalues is zero and the signals are linearly dependent. When the phase angle is $\pi/2$, it is possible to find the principal axes only if the mean values have been included in the correlation matrix, because the only extra-diagonal term consists of their product. If this term is zero, the matrix is already diagonal and the principal axes are the original axes. Moreover, the correlation matrix obtained with non-Gaussian signals does not generally have such a dominant diagonal as in the case of normal signals. In the limit, the principal component analysis might become meaningless.

Of course, such academic cases do not occur in practice due to the high dimension of the data set including noise: but, nevertheless, this simple example shows the specific behaviour of harmonic signals by comparison with normally distributed ones.

For the rectangular cylinder, the POD carried out on all the pressure distribution, including the mean value, can deliver a physical meaning of the proper functions only by analysing also the time evolution of the principal components, with Fourier analysis for instance. This was performed by Hémon & Santi (2002). However, it is possible to extract the mean value of the POD, and the physical analysis needs also to be carried out on both the proper function and the principal components. The two calculations are in fact similar, since heterogeneity of mean values always distorts the proper functions.

In order to eliminate the distortion one may carry out the POD by parts, by analysing clusters of data having more or less the same mean value and the same physical significance: for instance the POD on the rectangular section could be performed three times, one for the front side, another for the rear side, and finally for the two lateral sides. However, the problem of cutting the clusters becomes critical for structures with more complex shapes and a general method of decomposition is needed to ensure robustness of the technique.

## 3. THE BIORTHOGONAL DECOMPOSITION

### *3.1. THE CLASSICAL BIORTHOGONAL DECOMPOSITION*

#### *3.1.1. General presentation*

The biorthogonal decomposition has been introduced by Aubry *et al*. (1991) and the rigorous mathematical formulation can be found in that paper. We present here a brief summary of the main results. The idea is to carry out a deterministic decomposition of a space-time signal without assuming other properties of this signal beyond its square-integrability. In practice, the signal will be the result of aerodynamic measurements or computations, pressure, velocity or any other quantity, which guarantees the above property.

The biorthogonal decomposition of a signal $\boldsymbol{U}(\boldsymbol{x},t)$ function of space $\boldsymbol{x}\in\Re^3$ and time $t\in\Re$, with $\boldsymbol{U}(\boldsymbol{x},t)\in L^2(\boldsymbol{X}\times T)$, $\boldsymbol{X}\subset\Re^3$ and $T\subset\Re$, is formally written as

$$\boldsymbol{U}(\boldsymbol{x},t)=\sum_{k=1}^{\infty}\alpha_k\ \psi_k(t)\ \varphi_k(\boldsymbol{x}). \qquad (6)$$

The biorthogonal decomposition theorem proves that decomposition (6) exists, converges in norm and that

$$\begin{aligned} &\alpha_1 \geq \alpha_2 \geq \ldots \geq 0, \\ &\lim_{M\to\infty} \alpha_M = 0, \\ &\langle \varphi_k, \varphi_l \rangle = \overline{\psi_k\ \psi_l} = \delta_{k,l}\ . \end{aligned} \tag{7}$$

Aubry *et al.* have called *topos* the spatial modes $\varphi_k(\boldsymbol{x})$ with $\varphi_k \in L^2(\boldsymbol{X})$ and *chronos* the temporal modes $\psi_k(t)$ with $\psi_k \in L^2(T)$. They proved that the topos, associated to the set of the eigenvalues $\alpha_k^2 = \lambda_k$ are the eigenmodes of the spatial correlation operator

$$\boldsymbol{Sc}(\boldsymbol{x},\boldsymbol{x}') = \int_T \boldsymbol{U}(\boldsymbol{x},t)\boldsymbol{U}^*(\boldsymbol{x}',t)\,dt\ , \tag{8}$$

which is Hermitian and nonnegative definite. The notation $U^*$ means the conjugate of $U$ in the general case of a complex signal. Simultaneously, the chronos associated to the same set of eigenvalues $\lambda_k$ are the eigenmodes of the temporal correlation operator

$$\boldsymbol{Tc}(t,t') = \int_X \boldsymbol{U}(\boldsymbol{x},t)\boldsymbol{U}^*(\boldsymbol{x},t')\,d\boldsymbol{x}\,. \tag{9}$$

What is remarkable is the fact that the eigenvalues $\alpha_k^2$ are common to topos and chronos, which was proved by using notably the symmetry property of the correlation operators. This means that chronos and topos are intrinsically coupled because they have the same eigenvalue. However, it is possible to separate the information, spatial and temporal, by multiplying them by the weight factor $\sqrt{\alpha_k}$ . This remark will be useful for the application of Section 5 on velocity field generation.

It can be shown that the continuous form of the BOD can be extended to the discrete form, in which case the correlation operators become the correlation matrices. This assumes, however, that the observation window of the signal is reasonably sufficient to be analysed, which means that the number of time steps and nodes is large enough. Aubry *et al.* (1991) have demonstrated the robustness of the BOD when additional samples (temporal or spatial) are taken into account in the original signal.

Another useful result in practice is the possibility to truncate decomposition (6) to *M* spatio-temporal structures and that the sum of remaining terms, i.e. the truncation error, is smaller than the first neglected eigenvalue.

It has also been shown that the global energy of the signal is equal the sum of the eigenvalues:

$$\iint_{X,T} \boldsymbol{U}(\boldsymbol{x},t)\,\boldsymbol{U}^*(\boldsymbol{x},t)\,d\boldsymbol{x}\,dt = \sum_{k=1}^{\infty}\alpha_k^2 = \mathrm{Tr}(\boldsymbol{Sc}) = \mathrm{Tr}(\boldsymbol{Tc}). \tag{10}$$

The global entropy of the signal characterizing its degree of disorder is defined starting from the relative eigenvalues $\lambda_{rk}$ given by

$$\lambda_{rk} = \lambda_k \Big/ \sum_{k=1}^{\infty}\lambda_k \ . \tag{11}$$

The expression of the global entropy is

$$H = -\lim_{M\to\infty}\frac{1}{\log M}\sum_{k=1}^{M}\lambda_{rk}\,\log\lambda_{rk} \ . \tag{12}$$

Due to the normalizing factor, the entropy $H$ can be compared for different signals and it characterises the disorder level of the signal. When all the energy is concentrated on the first term of the decomposition, the entropy is zero. When the energy is equally distributed among all the terms, the entropy reaches its maximum, equal to 1. In particular, the entropy is a useful indicator when studying signals with respect to an external parameter. Aubry *et al.* (1991) presented the entropy as a good way to detect the transition to turbulence for instance.

#### *3.1.2. Links between BOD and POD*

According to the developers of the biorthogonal decomposition (BOD) themselves, there is no real link between BOD and POD, since they are based on fundamentally different principles. In fact, BOD can be seen as a time-space symmetric version of the Karhunen-Loève expansion or, in other words, a combination of the classical POD and the snapshot POD. However, the main difference of concern for our problem is the assumptions on the analysed signal, which has to be square-integrable only for the BOD, instead of square-integrable, ergodic and stationary for the POD. The biorthogonal decomposition is a more general method and the proper orthogonal decomposition method should be considered as a particular case.

Moreover, the BOD is not derived from an optimization problem of the mean square projection of the signal as in POD, although the method of calculation of BOD leads also to an eigenvalue problem of a correlation operator. The geometrical interpretation in state space, especially the principal axes of the ellipsoid vanishes in the case of BOD.

The main consequence of this concerns the discussion regarding the mean value which has to be kept in the decomposition: there is indeed a problem in choosing among the temporal mean, the spatial mean or even the global mean. Moreover, Aubry *et al.* (1991) showed that for instance in the decomposition of the temporally centred signal

$$\boldsymbol{U}(\boldsymbol{x},t)-\overline{\boldsymbol{U}(\boldsymbol{x})}=\sum_{k=1}^{\infty}\alpha_k\ \left(\psi_k(t)-\overline{\psi_k}\right)\ \varphi_k(\boldsymbol{x}). \qquad (13)$$

The centred chronos $\psi_k(t)-\overline{\psi_k}$ are not orthogonal in $L^2(T)$. In this case the eigenvalues are multiplied by the factor $\sqrt{1-\overline{\psi_k}^2}$. The remark applies similarly to a spatially centred signal. To reinforce the discussion, Aubry *et al.* (1991) recall that excluding "*the mean of the topos is equivalent to introducing an artificial correlation among them equal to* $-\langle\varphi_k\rangle\langle\varphi_l\rangle$ *for all pairs* $(\varphi_k,\varphi_l)$*. Similarly, centring the chronos introduces a correlation* $-\overline{\psi_k}\ \overline{\psi_l}$ ".

### 3.2. APPLICATION TO WALL-PRESSURES

The method of biorthogonal decomposition presented above is general and can be applied to any kind of spatio-temporal signals. In this section the technique is adapted to wall pressure distribution and the computation procedure is given. An extension is then carried out in order to analyse the aerodynamic force decomposition.

In carrying out the BOD, it is first necessary to choose one kind of correlation, spatial or temporal, and to find its eigenmodes. Secondly, the missing component, i.e. the chronos or the topos respectively, is computed with the help of the decomposition in equation (6) and a suitable normalisation. The calculation procedure is in fact very close to the one of the direct POD or snapshot POD, completed by normalisation. The numerical algorithm used here is the subspace method of Bathe and Wilson but other standard eigenvalue solvers may be used.

In the standard case of pressure distribution decomposition, in relation to aerodynamic force analysis, it is obvious that the expressions derived for the lift force by Hémon & Santi (2002) are still valid, by letting the topos equal to the proper functions and the chronos equal to the principal components suitably normalized. However, there is the possibility to select *a priori* the aerodynamic component which is involved in the study: for instance, for the rectangular section which oscillates in a direction normal to the flow, the main interest is in the lift force.

We make the decomposition of the local lift force

$$C_z^i(t) = p_i(t)\, A_i\, n_z^i\,, \tag{14}$$

where $i$ refers to the pressure tap number and $A_i$ is the elementary area. In what follows, the other components, drag and pitching moment, are easily obtained by extension. When the taps are not uniformly distributed, they have not the same weight and this should be corrected. Jeong *et al.* (2000) have proposed such a correction based on the weighting by the area element of the tap, and applied to classical POD. In the present case, a correction is useless, since we just modify the original pressure signal: the problem of area estimation and its normal vector is supposed to be solved by a pre-processor of the BOD, depending only on the grid of pressure taps.

The terms of the spatial correlation matrix are given by

$$\boldsymbol{Sc}_{i,j} = \overline{p_i\, p_j}\; A_i\; A_j\; n_z^i\; n_z^j \tag{15}$$

and leads satisfactorily to a symmetric matrix. However, some terms can be nullified when the wall under consideration is parallel to the force direction which is under analysis. In practice it will be impossible to carry out the eigenmode calculation without restriction of the domain to non zero terms. Aubry *et al.* (1991) have verified that for a restricted domain in $L^2(\boldsymbol{X})$ (or $L^2(T)$), the restricted spatial correlation operator (or temporal) preserves the BOD properties. The elimination of the useless taps from the decomposition is possible: in practice it has to be done in order to avoid any numerical distortion in the eigenvalue problem resolution.

The interesting point is that the resulting force is now directly decomposed in chronos and topos since we have $C_z(t) = \left\langle C_z^i(t) \right\rangle$, assuming suitable normalisation of the reference area. The BOD is now written as

$$C_z(t) = \sum_{k=1}^{M} C_{zk}(t) = \sum_{k=1}^{M} \alpha_k\, \psi_k(t) \langle \varphi_k \rangle\,. \tag{16}$$

It can be shown that the properties of orthogonality are preserved for the quadratic mean values and lead to an indicator of convergence of the decomposition, in terms of force:

$$E_{zk} = \overline{C_{z_k}(t)^2} \Big/ \overline{C_z(t)^2}\,, \tag{17}$$

which is usually given as a percentage. An entropy, specifically dedicated to the lift force, can then be defined by analogy:

$$H_z = -\frac{1}{\log M}\sum_{k=1}^{M} E_{zk}\,\log E_{zk} \quad . \tag{18}$$

In order to illustrate the application of these tools, we present in the following section two examples on classical bluff body flows. It should be noted before that the simultaneous analysis of two components, for instance X and Z, is mathematically possible although it has no sense from a mechanical point of view, the two components being already orthogonal in the physical space. In this case, the phase difference between the two fluctuating forces is fixed by the chronos of each component independently.

## 4. APPLICATION OF THE BOD ON PRESSURE DISTRIBUTIONS

### *4.1. RECTANGULAR PRISM*

This case was presented in detail by Hémon & Santi (2002) and we recall here the main points. The pressure data are the results from the computations with a Navier-Stokes solver. It solves the 2 D incompressible unsteady equations without turbulence modelling. The section of the rectangle has a length-to-thickness ratio of 2, and the Reynolds number is 6000 based on the length. Grid refinement was chosen so as to capture correctly the physics of the large coherent structures.

The main mechanism is a leading-edge alternate vortex shedding generating a quasi-periodic lift force. However, the pressure distribution along the lateral side is not uniform even in its unsteady part, due to the unsteady impingement of secondary vortices inside the mean shear layer, as shown in Figure 4.

#### *4.1.1. Inclusion of the mean component*

We shortly discuss in this part the inclusion of the mean component of the pressure. The first three topos are given in Figure 5, with inclusion of the mean on the left side, and with exclusion of the time mean component on the right side. The decompositions in the *X*- and *Z*-directions are given, following the procedure of the previous section, the two calculations being carried out independently. Case 1a represents obviously the mean pressure distribution.

For the *Z*-component, it can be seen that the next topos, 2a and 3a, are similar (eventually up to an arbitrary sign) to 1b and 3b, respectively, which are the first two topos related to the fluctuating part. It is interesting to see that a trivial interpretation in the state space as for the POD is not valid here, although the mean pressure is more or less homogeneous on the lateral sides. Keeping the mean value does not perturb the topos in this case. Moreover, in case 2b there appears to be a mean pressure. It was however removed and cannot represent correctly the physical pressure distribution.

Concerning the *X*-decomposition, it is fundamentally different from its *Z* counterpart due to the heterogeneous mean values. We can see that Case 2a is similar to 1b. However, Case 3a has no equivalent when the mean value is excluded, although it has a physical significance from the aerodynamic point of view (out of phase fluctuation of the front and rear sides, leading to partial neutralization of the fluctuating drag at the fundamental Strouhal number). Symmetrically Case 2b has no equivalent and is limited to the rear side for its physical interpretation. Additionally Case 3b has an equivalent which is not shown (it would be 4a).

In summary, one should conclude here that exclusion of the time-mean component is not a method which simplifies the BOD: when it is included, the physical meaning is easier to find because it is ensured that the first term of the decomposition is a representative of the mean and that the higher order terms cannot be perturbed by the mean because of the orthogonality property.

*4.1.2. BOD results*

The results of the BOD are given in Figures 6 (eigenvalues), 7 (entropies), 8 and 9 (topos), and 10 (chronos). The three decompositions in the *X*- and *Z*-directions and pitching moment were calculated. An additional case was also done, which consists in forcing a periodic motion of the cylinder in the *Z*-direction with a frequency corresponding to a galloping regime, see (Hémon & Santi 2002) for a detailed presentation. Four terms of the decomposition are reasonably sufficient to represent the main mechanism.

It is shown from the computed eigenvalues that the BOD converges rapidly. The drag converges faster because most of its energy is included in the mean. It is clearer with the global entropies which are minimum for drag, and reinforced for the specific drag entropy ($H_x$). The effect of the motion on the lift is to increase the related entropy, mainly through the second term because the energy (the eigenvalue) taken by this term is larger when the cylinder is in motion (Figure 6). It seems logical from a mechanical point of view that a forced motion brings some energy in the system, leading to a higher entropy.

One should notice also the smaller value of the specific entropies ($H_x$ or $H_z$) by comparison with the global entropies. This difference expresses well the efficiency of the decomposition carried out on local forces as proposed in Section 3.2, since in the limit, zero entropy would mean that only the first term contains the complete force in the $L^2(T)$ sense.

Concerning the topos and the chronos (Figures 8, 9 and 10), we must recall that the sign is not significant because it is their product which has a physical meaning. The negative mean value of the first chronos in the *X*-direction has to be multiplied by its negative counterpart in the corresponding topos, thereby keeping the product positive as one would expect for the mean drag coefficient.

The analysis of the *Z*-component is more significant in this problem. We observe that the second term of the BOD is the fluctuating force generated by the leading edge vortex shedding embracing all the lateral sides. The third term, which creates the main part of the fluctuating pitching moment, is concerned with the periodic impinging of the main leading edge vortices. At a smaller scale, the fourth term is the first one related to the secondary vortices inside the shear layer: this weaker mechanism requires however higher terms to be correctly described.

The analysis of the pitching moment with respect to the centre of the cylinder was carried out using only the local forces in the *Z*-direction, the contribution of the local drag forces being neglected. The topos are shown in Figure 9, and the chronos were found similar to those of the lift force. It is interesting to notice that the topos are in fact exactly similar also, the difference being due to the product of the local lift by the distance to the centre of the cylinder. Indeed, the local pitching moment is

$$C_M^i(t) = p_i(t)\, A_i\, d_O^i\, n_z^i\,, \qquad (19)$$

where $d_O^i$ is the projection on the x-axis of the distance from node $i$ to the centre of gravity $O$ of the section. The corresponding diagonal matrix with diagonal elements $d_O^i$ is denoted $\boldsymbol{D}_O$. The topos of the BOD for the pitching moment are the eigenmodes of the related spatial correlation matrix which can be written as

$$\boldsymbol{R}' = \boldsymbol{D}_O\, \boldsymbol{R}\, \boldsymbol{D}_O\,, \qquad (20)$$

where $\boldsymbol{R}$ is the spatial correlation matrix for the local lift. It can be shown that the topos of the pitching moment are exactly those of the lift multiplied by the distance to the centre of gravity, i.e.

$$\varphi'_k = \boldsymbol{D}_O \; \varphi_k \; . \qquad (21)$$

The new associated eigenvalues can be expressed also in terms of those of the lift decomposition as

$$\lambda'_k = \lambda_k \left\langle \boldsymbol{D}_O \; \varphi_k , \boldsymbol{D}_O \; \varphi_k \right\rangle . \qquad (22)$$

The transformation which makes the lift become a pitching moment is a linear geometrical operation independent of time and therefore it is directly transmitted on the topos only.

In conclusion, for this example, it should be said that the biorthogonal decomposition of the local force can be a useful tool for (i) understanding the different mechanisms involved in the aerodynamic total load, (ii) making the separation into simple components, and (iii) obviously reducing the amount of data to be stored. As long as the Fourier spectrum of the chronos is pure, i.e. with a few discrete frequencies, the use of the adequate terms of the BOD in a structural response computation tool, as suggested by Dowell *et al.* (1999) becomes possible.

### *4.2. CIRCULAR CYLINDER AT LOW REYNOLDS NUMBER*

As the previous example was designed with straight walls, only parallel or normal to the directions of the drag or lift, the present circular cylinder is different. Its characteristic geometry leads to a progressive continuous mean pressure distribution which does not create well-formed clusters of data in state space as for the rectangle. This illustration case is therefore complementary to the previous one.

The 2 D circular cylinder is one of the most documented cases of bluff body flow due to the alternate vortex street generated for a given range of Reynolds numbers. The purpose of this section is limited to the illustration of the BOD method: the pressure data are obtained through a numerical simulation of the flow using the same Navier-Stokes solver as for the rectangular section. The simulated equivalent Reynolds number is 150, leading to a well established alternate vortex regime. The data-set consists of 100 nodes with 950 time steps, representing 19 periods of the alternate shedding. Three kinds of BOD have been calculated, with the pressure distribution (scalar) and with the local forces in the longitudinal (*X*) and lateral (*Z*) directions.

The cumulated entropies are given in Figure 11. For the BOD on the pressure, the specific entropies obtained by recombination of drag and lift are given also: they are not equivalent to those obtained by decomposition of local forces. The specific entropies show the large difference between the *X*- and *Z*-directions: the method based on the *X*- or *Z*-decomposition of the local force is seen to be much sensitive to the selected direction than the corresponding decomposition of the pressure. This demonstrates objectively that the spatio-temporal complexity is much larger for the lift component (two orders of magnitude) than for the drag component. One must recall however that this criterion is based on mean-square values.

The topos for the pressure distribution are given in Figures 12 and 13 for the local forces. The chronos are given in Figure 14 for which only one set is presented since all three were found similar, with differences only in their sign, in accordance with the sign of the corresponding topos.

In each case, the first structure is the mean component, the second and third ones the main fluctuating components at the fundamental Strouhal number and the fourth one for twice the Strouhal number, as can be seen on the chronos. It is interesting to see that the decomposition of local forces seems richer than that for the pressure, although the projection of the topos of the pressure distribution in the *X*- or *Z*-directions leads to similar topos. Nevertheless, the local force decomposition is easier to analyse notably because the sensitivity is increased, and that better separation is achieved.

For instance for the lift, the second structure corresponds to the unsteady stall embracing all the surface of the cylinder on the same side with the same sign of the pressure. The third structure, with a phase lag of $\pi/2$ in time with respect to the second, corresponds to a local force distribution which expresses the out of phase behaviour between the front and rear part of one side of the cylinder.

Considering only the structures 2 and 3, one could object here that the orthogonality requirements should obviously lead to such topos and chronos. This has to be tempered (i) by the fact that the physics can lead here to such results because the signals are relatively close to being harmonic, and (ii) by the small eigenvalue affected in the third structure which makes it not so important in amplitude.

In summary, this simple example confirms the previous conclusions obtained with the rectangular cylinder and adds a complementary remark concerning the separation *a priori* of the force components, or *a posteriori* by recombination with the BOD of the pressure: in such a simple case, the geometry does not lead to clusters in state space, and separation of components *a priori* is not so efficient. This is reinforced by the fact that the chronos were found identical in the three BOD. Nevertheless, for complex structures, involving sharp angles for instance, the spatial complexity is such that the separation *a priori* of the components will lead to better efficiency of the decomposition.

## 5. SIMULATION OF A SPATIALLY CORRELATED VELOCITY FIELD

We present in this section an application of the BOD to the modelling of a velocity field. The purpose is very different from the previous sections where the BOD was used as an analysis tool for already available signals: the objective here is to generate the signal by using the specific properties of the BOD.

In the field of wind-excited structures, temporal simulations are increasingly important, due to the large size of the structures as for instance in modern suspended bridges (AFGC, 2002). As a consequence, there is a need for more accurate computations which can include the nonlinear behaviour of the structure. However, the classical techniques for solving the dynamical problem are generally based on spectral methods in which the nonlinear part, structural and even aeroelastic, is difficult to introduce.

The temporal simulation of the aeroelastic coupled problem, including atmospheric turbulence excitation, requires considerable input data in terms of a velocity field: this field has to be close to the real turbulent wind, one essential characteristic of which is the spatial correlation. There exist a number of methods for generating a turbulent velocity field as presented in the review by Guillin & Crémona (1997) and Di Paola (1998).

One of them is derived from the method proposed by Yamazaki & Shinozuka (1990) for application in earthquake engineering. Their approach which is called *statistical preconditioning,* is based on the modal decomposition of the spatial covariance matrix and the temporal part of the signal is generated by using a Fourier decomposition. Recently, Carassale & Solari (2002) similarly used the direct POD to generate a turbulent wind velocity field and to compute the wind loads acting on the eigenmodes of a structure.

The method can be improved by exploiting the space-time symmetry of the BOD, as outlined below. We emphasise that our objective is to illustrate the use of the BOD technique. We will not enter into the details of the physical modelling of the wind: more information can be found in the related literature, as in (AFGC, 2002) and references therein.

### *5.1. Initial Data*

In the civil engineering community, the turbulent wind is usually described by a few statistical parameters which constitute the targets for the simulated wind field. We derive in this section the expressions related to the vertical velocity component $w$ of the turbulence which is applied to an elongated horizontal structure, such as a bridge deck. The formulation can readily be extended to the other components.

The target parameters describing the atmospheric turbulent wind are (i) the horizontal mean velocity $\bar{U}(z)$, which may be a function of the altitude $z$, (ii) the standard deviation of the vertical velocity $\sigma_w$, (iii) the Power Spectral Density of the vertical velocity $S_w(f)$ versus frequency $f$, and (iv) the coherence function of the vertical velocity in the lateral direction $\gamma_w^y(f)$.

The power spectral density function and the coherence function may come from different sources and we have chosen here the most common modelling, i.e. the von Kármán spectrum

$$\frac{S_w(f)}{\sigma_w^2} = \frac{4\,L_w^x}{\bar{U}(z)} \frac{1+188.4\left(\dfrac{2\,f\,L_w^x}{\bar{U}(z)}\right)^2}{\left(1+70.7\left(\dfrac{2\,f\,L_w^x}{\bar{U}(z)}\right)^2\right)^{11/6}}, \qquad (23)$$

and an exponential coherence function between points $i$ and $j$ given by

$$\gamma_w^{i,j}(f) = \exp\left[\frac{-C_w^y\,|y_i - y_j|\,f}{\bar{U}(z)}\right], \qquad (24)$$

where $L_w^x$ is the longitudinal scale of the vertical velocity component and $C_w^y$ the coherence coefficient of the vertical velocity in the lateral $y$ direction. It should be noted that the power spectral density is normalized by definition with the square of the standard deviation.

All the parameters appearing in the target characteristics are usually extracted from literature data or from *in situ* measurements and meteorological studies, especially when the site effect can be significant, as for instance in a mountainous area.

It should be noted that a complete 2 D field cannot be built due to the lack of modelling for the cross-spectral density function: for instance for a horizontal and vertical structure, such as a suspended bridge including the deck and the pylons, the cross-spectrum between lateral and vertical turbulence is needed. Unfortunately, there is actually no generally accepted modelling. Therefore, we give in the following a single-component application that may be extended to two components, provided the cross-spectrum is known.

### 5.2. Development of the Method with BOD

The idea is to generate a wind velocity field which is given by a BOD as

$$\boldsymbol{U}(y,t)=\sum_{m=1}^{M}\sqrt{\alpha_m^t}\;\psi_m(t)\;\sqrt{\alpha_m^y}\;\varphi_m(y), \tag{25}$$

where the chronos are associated with the set of eigenvalues $\left(\alpha_m^t\right)^2$ and the topos with $\left(\alpha_m^y\right)^2$. The main point is to find the topos and the chronos separately by solving twice the corresponding eigenvalue problem.

By assuming a suitable discretisation in time and space, which will be discussed below, the spatial correlation matrix is built starting from the spectral density functions between nodes $i$ and $j$ as

$$\boldsymbol{Sc}_{i,j}=\sum_l\sqrt{S_{w_i}(f_l)\,S_{w_j}(f_l)}\;\gamma_w^{i,j}(f_l), \tag{26}$$

where the index $l$ refers to the frequency. The integration over the frequency band has to be compatible with the time discretisation.

For the chronos, we assume that the individual signals at point $i$ and time $t_k$ are built with Fourier series as

$$w_i(t_k)=\sum_l\sqrt{2\,S_{w_i}(f_l)}\;\cos\!\left(2\pi\;f_l\;t_k+\phi_{i,l}\right), \tag{27}$$

where the phase angles $\phi_{i,l}$ are randomly uniformly distributed in $[0,2\pi]$. The temporal correlation matrix is given by

$$\boldsymbol{Tc}_{k,n}=\left\langle w_i(t_k)\,w_i(t_n)\right\rangle . \tag{28}$$

From the definitions given above, it can be shown that the total energy of the signal given by the trace of each correlation matrix is

$$\mathrm{Tr}(\boldsymbol{Tc})=K\;\sigma_w^2\;,\quad \mathrm{Tr}(\boldsymbol{Sc})=N\;\sigma_w^2\;, \tag{29}$$

where $K$ and $N$ are the number of time steps and the number of nodes respectively. It becomes therefore obvious that these numbers have to be equal, a constraint which did not appear clearly in the previous section. The time-space symmetry of the BOD was not really employed for the analysis of existing available signals, and the dimensions of each set, temporal or spatial, were implicitly assumed to be equal. In the present case, the explicit equality is necessary and the following application will illustrate the constraints that result from this equality.

### 5.3. APPLICATION TO A BRIDGE DECK

In this example, the objective is to generate a turbulent vertical velocity signal applied to a horizontal bridge deck. The atmospheric turbulence characteristics are given in Table 1 and the simulation characteristics in Table 2. These data are realistic and correspond to one configuration of the Millau bridge French project.

In carrying the temporal simulation, it is obvious that the temporal window has to be compatible with the frequency band required. On the one hand, the time step is imposed by the highest frequency to be reproduced, through the Shannon sampling theorem. On the other hand, the total duration of the simulation should be compatible with the lowest frequency requirement. In practice for such civil engineering structures, the turbulence spectrum and the structural spectrum, are concentrated in the lower most energetic frequencies, corresponding to the large turbulence scales and to the length of the bridge. From the frequency band of Table 2, the time step was chosen equal to 1/6 s with 256 points, leading to a duration of 42.7 s. The latter is just greater than 2 periods of the lowest frequency. The spatial resolution is consequently 1.33 m.

A few results of the velocity field generation are presented in Figures 15 (samples), 16 (power spectral densities) and 17 (spatial correlation). The four samples of Figure 15 have a standard deviation of 4.32, 4.27, 4.28 and 4.28 respectively which must be compared to the target value of 4.38. The small deficit is due to the lower frequency range which is not well reproduced by the short duration of the simulation. By increasing the number of points, up to 512, the standard deviation deficit vanishes. The comparison between the target and the simulated power spectral densities in Figure 16 shows the satisfactory behaviour of the method, the lower frequencies being weakly reproduced for the reason outlined above.

The comparison of the target and the simulated spatial correlations, in Figure 17, is also satisfactory. The integration of these curves gives for the target $L_w^y$ = 40 m against 35 m for the simulation.

It is also interesting to check the convergence of the BOD, and the number of spatio-temporal structures that are needed in order to reach a certain level of accuracy. The indicators are the cumulated eigenvalues of the chronos and the topos which must be compared with the theoretical value, i.e. the variance. Figure 18 shows this comparison. We see that with less than 40 modes, more than 90% of the energy is included. The storage of data for the velocity field is considerably decreased since we need a size of 2x40x256 by comparison with 256x256 for a classical simulation method, giving rise to a factor larger than 3.

It should be said also, from a physical point of view, that the higher frequencies are spatially less correlated, which is a characteristic of the turbulence. Due to this, the convergence of the BOD is always better in the low frequency range.

The last remark concerns the signal construction by equation (25), with the help of topos and chronos that are computed independently. This is the central point of the technique. It implies that for a given chronos, the associated topos (i.e. of same order) is fixed and reciprocally. This assumption is not ensured *a priori* and should require further mathematical investigations. The results show however that the behaviour is very satisfactory because the two sets of eigenvalues (topos and chronos) have a very close convergence evolution, as seen in Figure 18.

Another point is the determinism of the method, although there is apparently a stochastic part in equation (27) for the chronos estimation. Complete determinism is obtained through the truncation of the decomposition by retaining the first $M$ chronos and topos. We give in Figure 19 the first four chronos and topos. We see that the topos have a wavelength which decreases as their order increases. Consequently, the associated chronos have a main frequency which increases simultaneously. It is consistent with the mechanical point of view and explains the good behaviour of the decomposition given by equation (25).

In conclusion, it should be said that the BOD is a very convenient, even elegant, technique for generating spatially correlated fields, such as atmospheric wind. It can be used also for earthquake engineering. Moreover, the reduction of the data to be stored, a critical point when performing temporal simulations, is significant and can be easily controlled.

## 6. CONCLUSION AND FURTHER APPLICATIONS

We have presented the biorthogonal decomposition of spatio-temporal signals proposed by Aubry *et al.* (1991) in the context of fluid-structure interaction problems. This tool generalizes the so called proper orthogonal decomposition in the sense that the signal is not necessarily ergodic and stationary. In most cases, the biorthogonal decomposition should replace the classical proper orthogonal decomposition and it should be applicable to a wider variety of situations.

For instance, we have adapted the method to the analysis of the wall pressure distribution and discussed the problem of the inclusion of the mean values, especially when heterogeneities are present. In such a case, we have shown that the POD is submitted to distortions because it is employed arbitrarily with signals which do not satisfy the fundamental assumptions.

We have also proposed to use the BOD in order to generate a turbulent velocity field: the method is based on the space-time symmetry of the BOD which is imperative for this application and not possible with POD. A test-case was presented and shows that the reduction of the amount of stored data is considerable by comparison with other methods based, for instance, on Fourier transform.

Various extensions can be done on the decomposed signals, for instance if the data are available, the decomposition onto local flutter derivatives could be performed, so that the different coherent structures detected could be associated to a specific structural action, such as aerodynamic damping. Amandolèse (2001) has partially initiated the process with POD. By using BOD instead, the specific entropy would be a very sensitive indicator in aeroelastic stability by reference to a control parameter such as the reduced velocity.

Still in the fluid-structure interactions domain, the BOD can be seen as a powerful method of filtering experimental data in order to make them more comprehensive and ready for suitable processing. A simple example is given by the problem of vortex-induced waves along cables currently studied at LadHyX (Facchinetti *et al.* 2002). In this application, a long flexible cable is submitted to vortex shedding excitation and the question is to detect the presence of travelling waves rather than standing ones. The experiments are carried out in a water tank at low Reynolds numbers, of the order of 100. The measurement technique involves a numerical video camera which provides a numerical movie. After processing, the lateral displacement of a given length of the cable is obtained versus time.

The BOD is carried out on this signal where the space dimension is reduced to a single vertical coordinate. The results show that only the first two terms are sufficient to recover correctly the original signal at 90% in energy, the first term having roughly an eigenvalue twice the second. The topos and chronos are given in Figure 20. The main frequency visible on the chronos allows to recover the Strouhal number. From these results, it is possible to demonstrate objectively the presence of travelling waves and also to quantify the incident and reflexive part by a quantitative use of the eigenvalues and of the phase lag between the chronos.

## APPENDIX: NOMENCLATURE

| | |
|---|---|
| $A_i$ | elementary area of node $i$ |
| $C_w^y$ | coherence coefficient of $w$ in the lateral direction |
| $C_z$ | global lift coefficient |
| $C_{z_n}$ | global lift coefficient contribution of order $n$ |
| $C_z^i$ | local lift coefficient of node $i$ |
| $E_{z_n}$ | relative contribution of order $n$ of the quadratic mean lift coefficient |
| $f$ | frequency |
| $H$ | in BOD, global entropy |
| $H_z$ | in BOD, entropy based on lift decomposition |
| $K$ | number of time steps |
| $L_w^x$ | longitudinal scale of $w$ |
| $M$ | number of modes |
| $N$ | number of pressure taps or nodes |
| $n_z^i$ | z component of the normal vector of node $i$ |
| $\boldsymbol{P}(\mathbf{x},t)$ | wall pressure distribution, function of space and time |
| $p_i(t)$ | pressure of node $i$, function of time |
| $\boldsymbol{R}$ | cross-correlation or covariance matrix |
| $\boldsymbol{Sc}$ | spatial correlation matrix |
| $S_{w_i}(f)$ | Power Spectral Density (PSD) of $w$ at point $i$ |
| $\boldsymbol{Tc}$ | temporal correlation matrix |
| $\boldsymbol{U}(\mathbf{x},t)$ | spatio-temporal signal, function of space and time |
| $\boldsymbol{W}_n(\boldsymbol{x})$ | in POD, proper function of order $n$, function of space |
| $w_i(t)$ | vertical velocity at point $i$, function of time |
| | |
| $\alpha_k$ | in BOD, coefficient of order $k$ , $\alpha_k = \sqrt{\lambda_k}$ |
| $\gamma_w^{i,j}(f)$ | coherence function of $w$ between nodes $i$ and $j$ |
| $\delta_{k,l}$ | Kronecker symbol |
| $\lambda_n$ | eigenvalue of order $n$ |
| $\lambda_{r\,n}$ | relative eigenvalue of order $n$ |
| $\mu_n(t)$ | in POD, principal component of order $n$, function of time |
| $\sigma_w$ | standard deviation of $w$ |
| $\varphi_k(\boldsymbol{x})$ | in BOD, topos of order $k$, function of space |
| $\phi$ | phase angle |
| $\psi_k(t)$ | in BOD, chronos of order $k$, function of time |
| | |
| $\bar{s}$ | denotes time integration of $s$ |
| $\langle s \rangle$ | denotes space integration of $s$ |
| $\langle r, s \rangle$ | denotes the Euclidian scalar product between $r$ and $s$ |

**TABLES**

*Table 1. Atmospheric turbulence characteristics*

*Table 2. Simulation parameters*

*Table 1*

| | |
|---|---|
| $\overline{U}$ | 36.55 m/s |
| $\sigma_w$ | 4.386 m/s |
| $\sigma_w / \overline{U}$ | 0.12 |
| $L_w^x$ | 30 m |
| $C_w^y$ | 9 |

*Table 2*

| | |
|---|---|
| Frequency band | 0.05 – 3 Hz |
| Length of the deck | 340 m |

## FIGURE CAPTIONS

Figure 1. Pressure taps number around the rectangular section.

Figure 2. State locus for 2 by 2 nearest nodes around the rectangular section of Figure 1.

Figure 3. Samples of pressure signals around the rectangular section. Left, pressure versus time (dimensionless) ; Right, corresponding probability density function.

Figure 4. Instantaneous vorticity distributions for the rectangular cylinder at Re=6000.

Figure 5. First three topos for the rectangular section with (1-3a) and without (1-3b) inclusion of the mean value. ▬▬, *Z*-direction ; ▬ ▬, *X*-direction.

Figure 6. Eigenvalues for the rectangular section: ○, *X*-direction ; Δ, *Z*-direction ; ▲, *Z*-direction in forced oscillations.

Figure 7. Cumulated entropy for the rectangular section: ○, *X*-direction ; Δ, *Z*-direction ; ▲, *Z*-direction in forced oscillations.

Figure 8. Topos of the rectangular section. ▬▬, *Z*-direction ; ▬ ▬, *X*-direction.

Figure 9. Topos of the rectangular section. Pitching moment in *Z* direction.

Figure 10. Chronos of the rectangular section. Left, *X*-direction ; Right, *Z*-direction.

Figure 11. Cumulated entropy for the circular cylinder. Filled symbols, pressure analysis ; Open symbols, local force analysis ; ○, ●, *X*-direction ; Δ, ▲, *Z*-direction..

Figure 12. Topos for the pressure distribution of the circular cylinder.

Figure 13. Topos of the circular cylinder. ▬▬, *Z*-direction ; ▬ ▬, *X*-direction.

Figure 14. Chronos of the circular cylinder.

Figure 15. Samples of generated time histories of $w(t)$.

Figure 16. Comparison of power spectral densities. ▬▬, target function ; ▬ ▬, samples of Figure 15.

Figure 17. Comparison of correlations. ▬▬, target function ; ▬ ▬, simulated

Figure 18. Convergence of the BOD: Δ, topos ; ○, chronos ; ▬▬, target $\sigma_w^2$ ; ▬ ▬, 90% of target

Figure 19. First four chronos (left) and topos (right) for the turbulent wind

Figure 20. First two chronos and topos of a vortex excited cable: ▬▬, first ; ▬ ▬, second

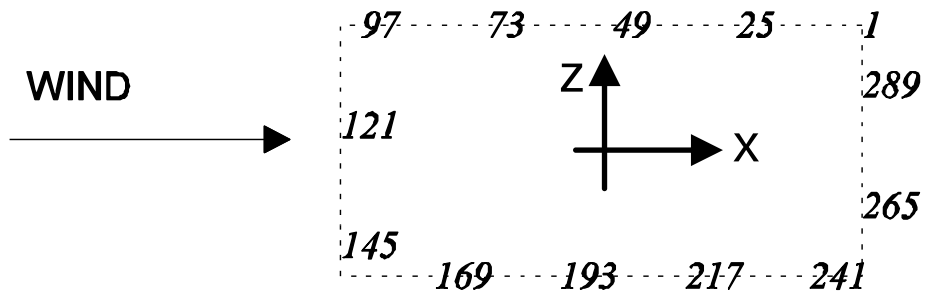

Figure 1.

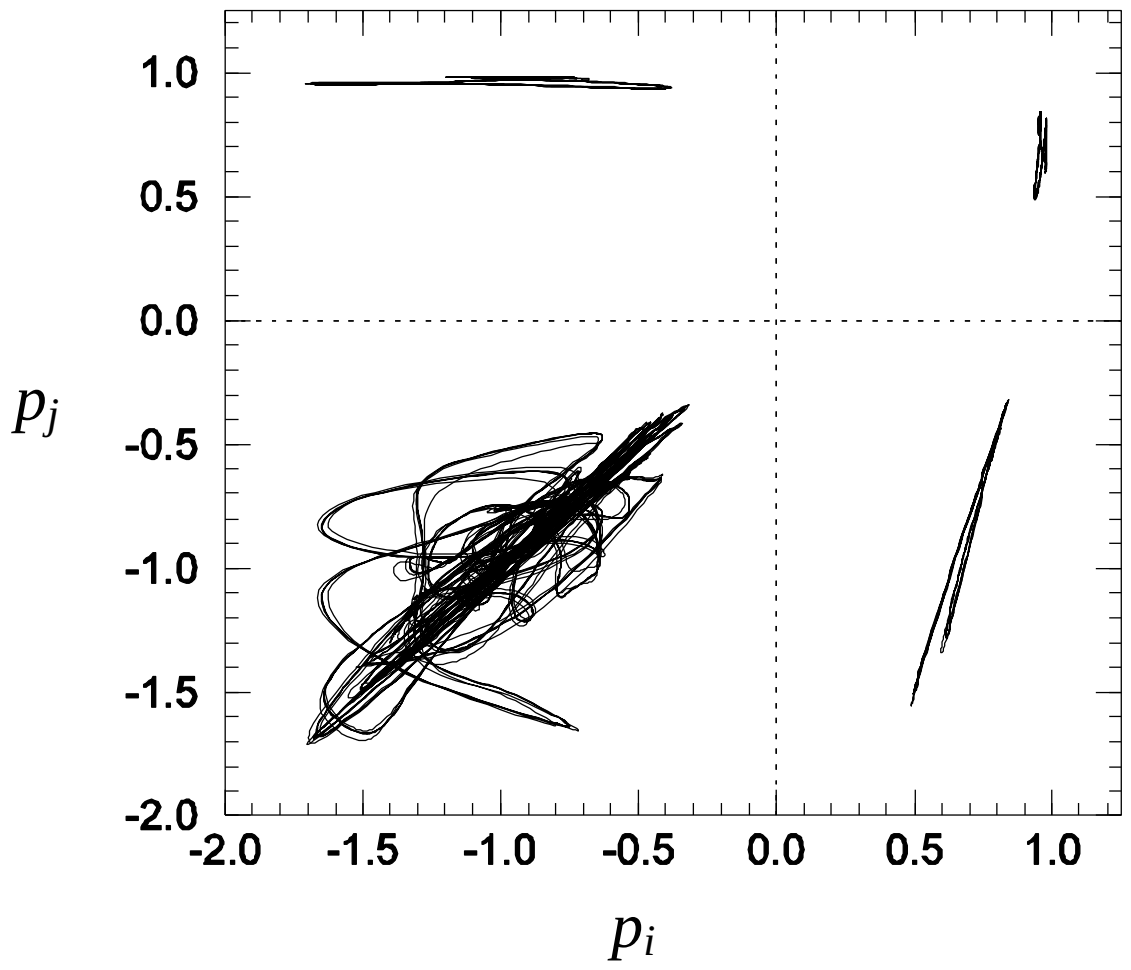

Figure 2.

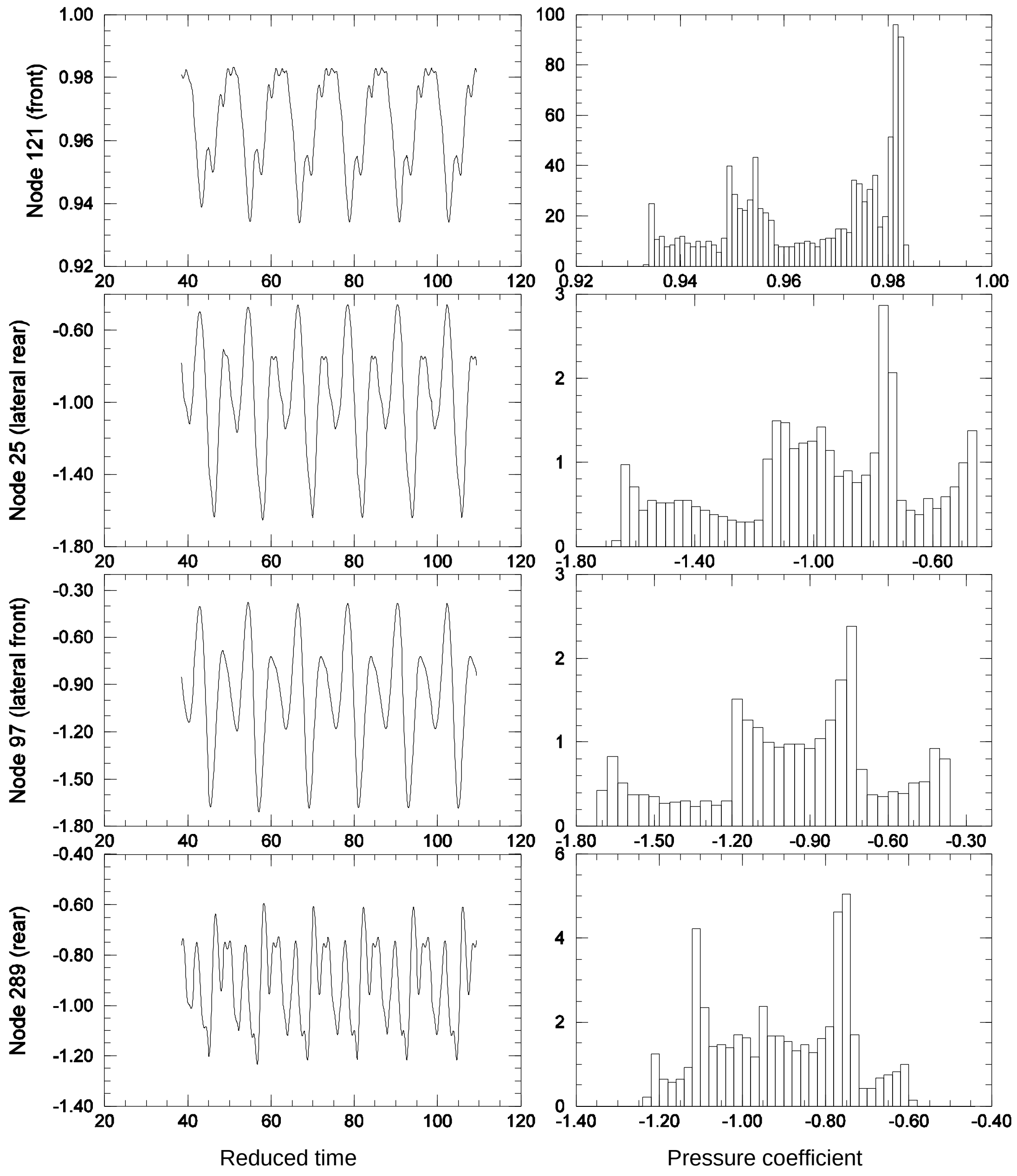
Node 121 (front)
Node 25 (lateral rear)
Node 97 (lateral front)
Node 289 (rear)
Reduced time
Pressure coefficient

Figure 3.

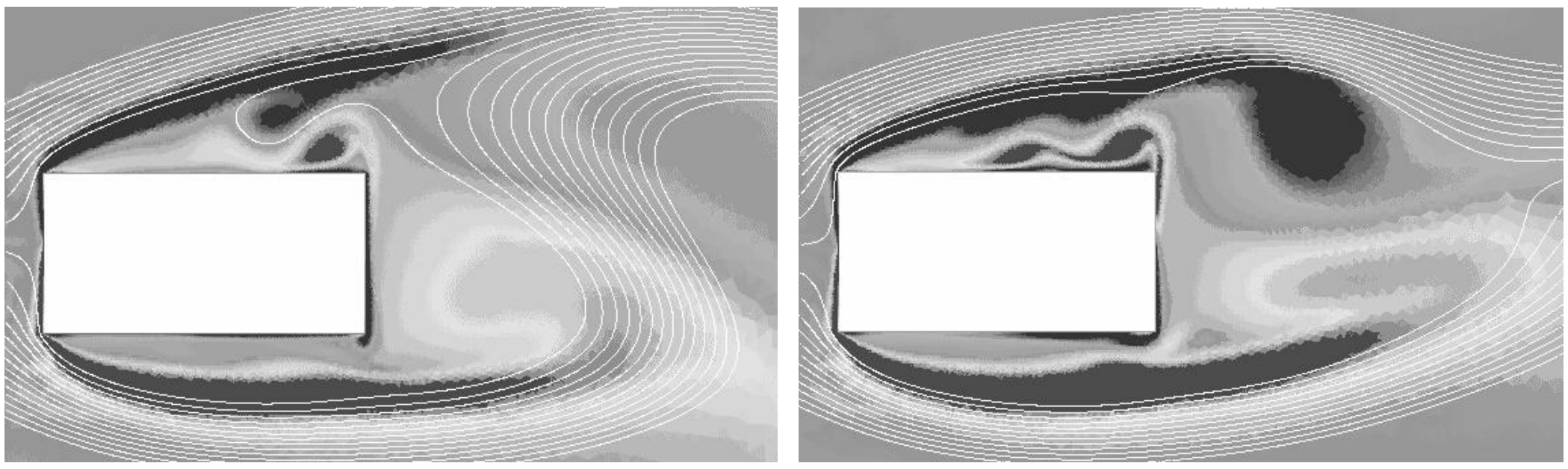

Figure 4.

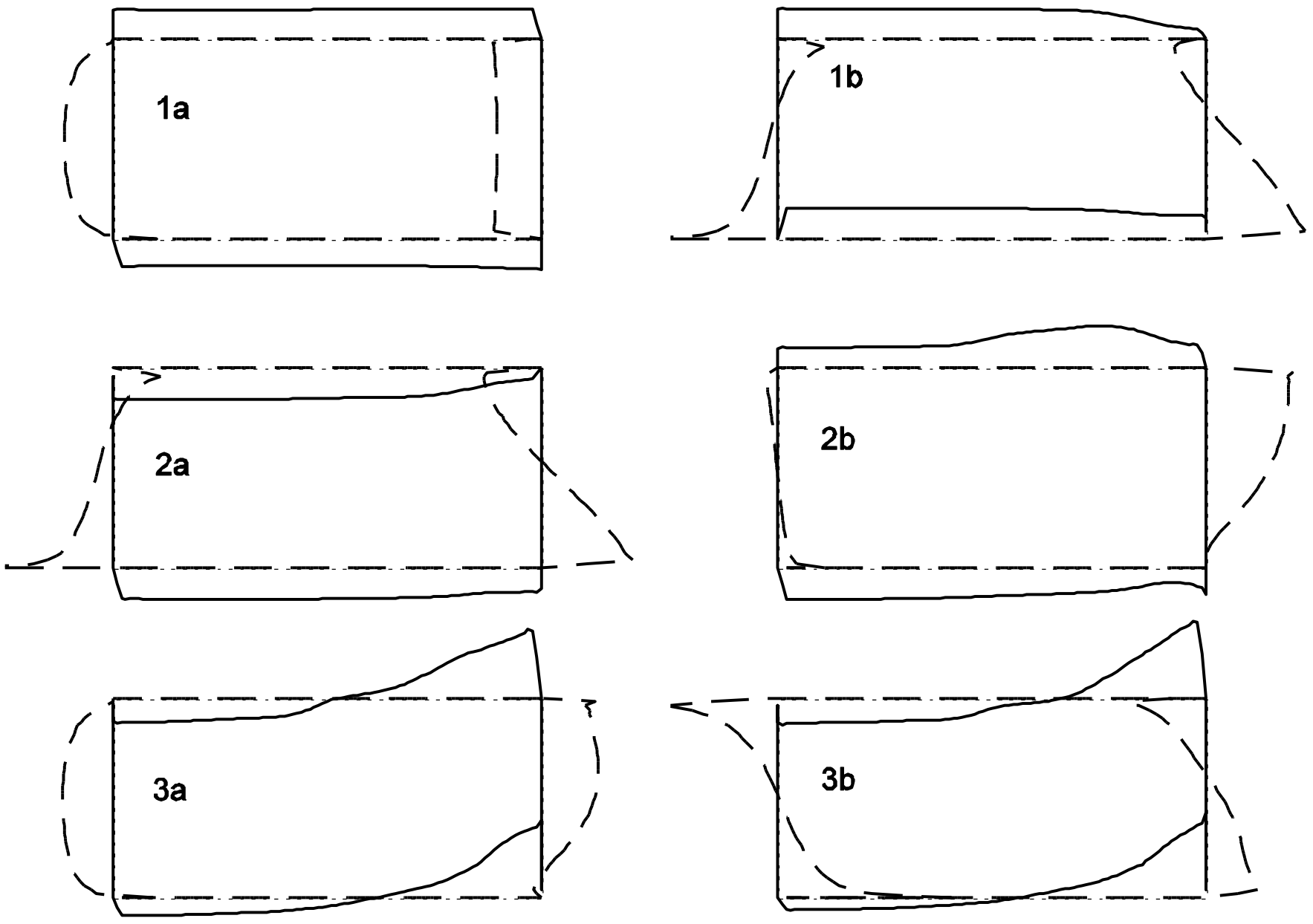

Figure 5.

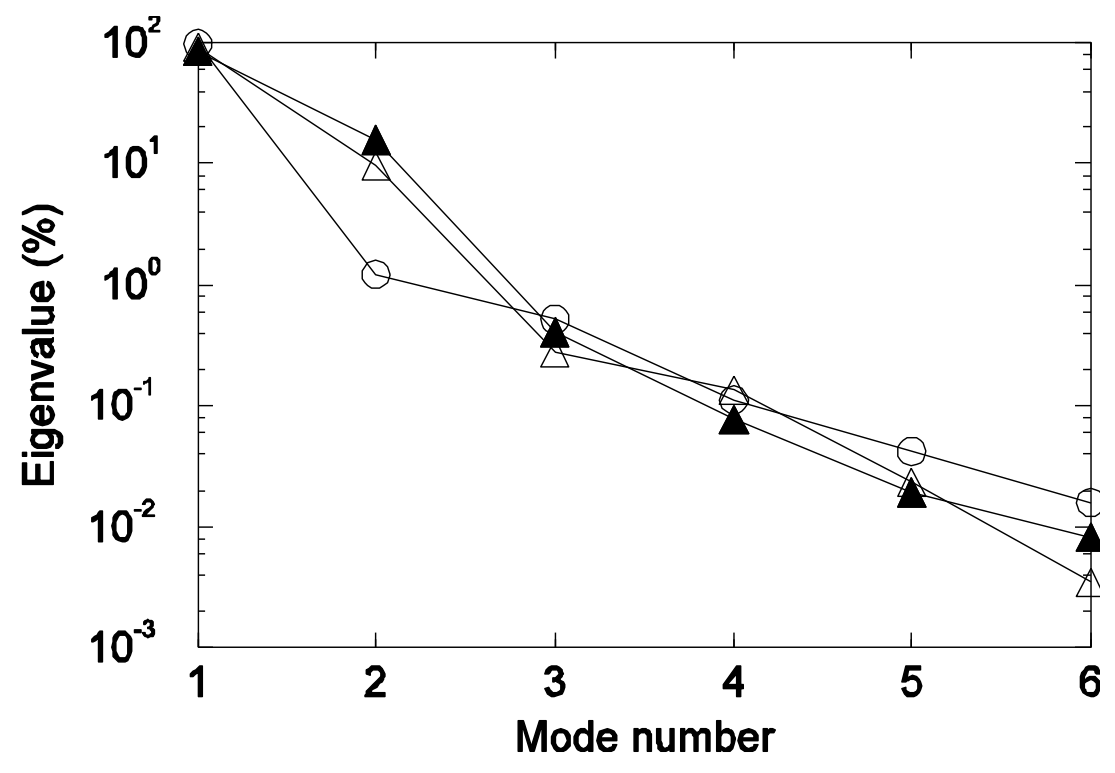

Figure 6.

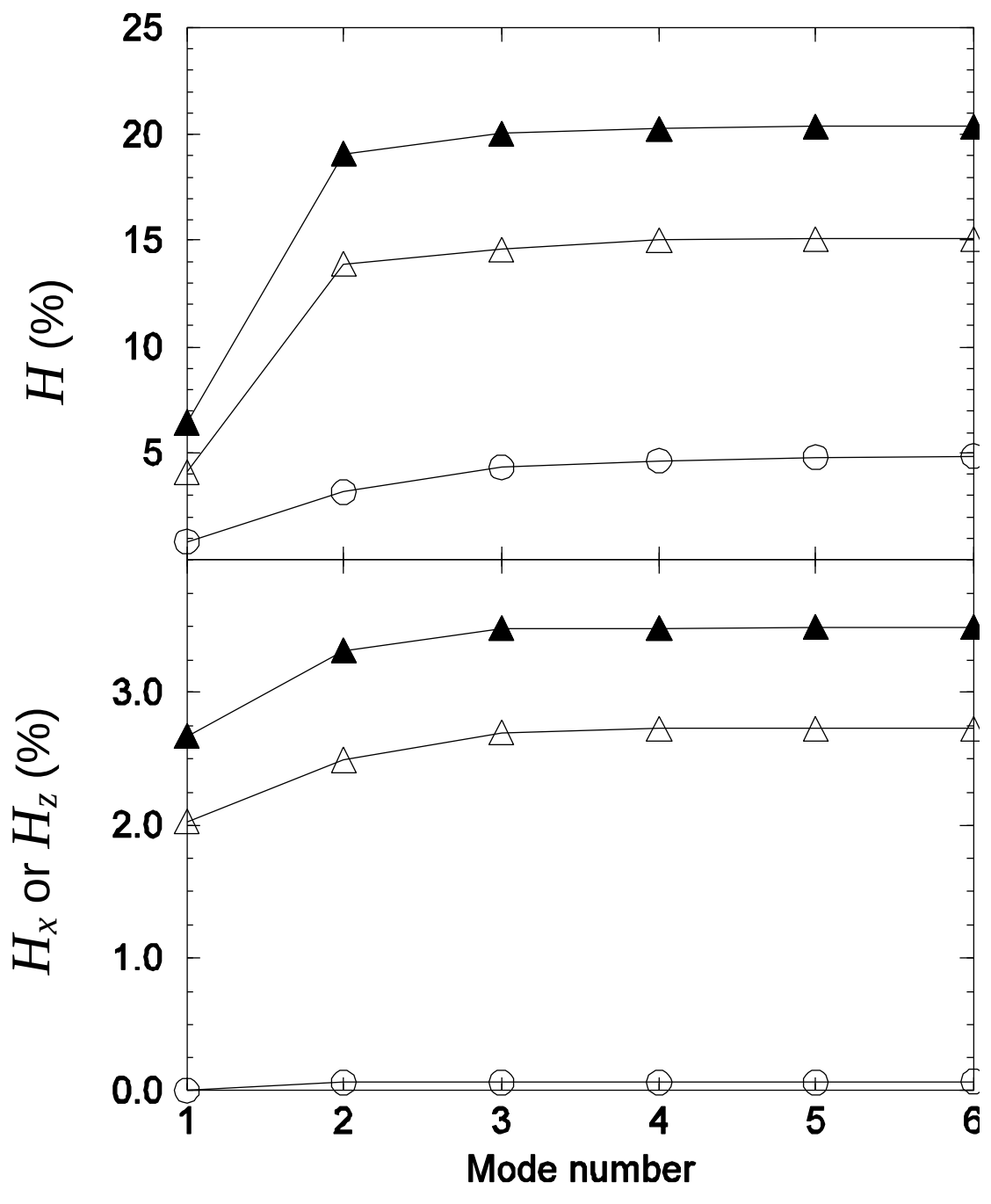

Figure 7.

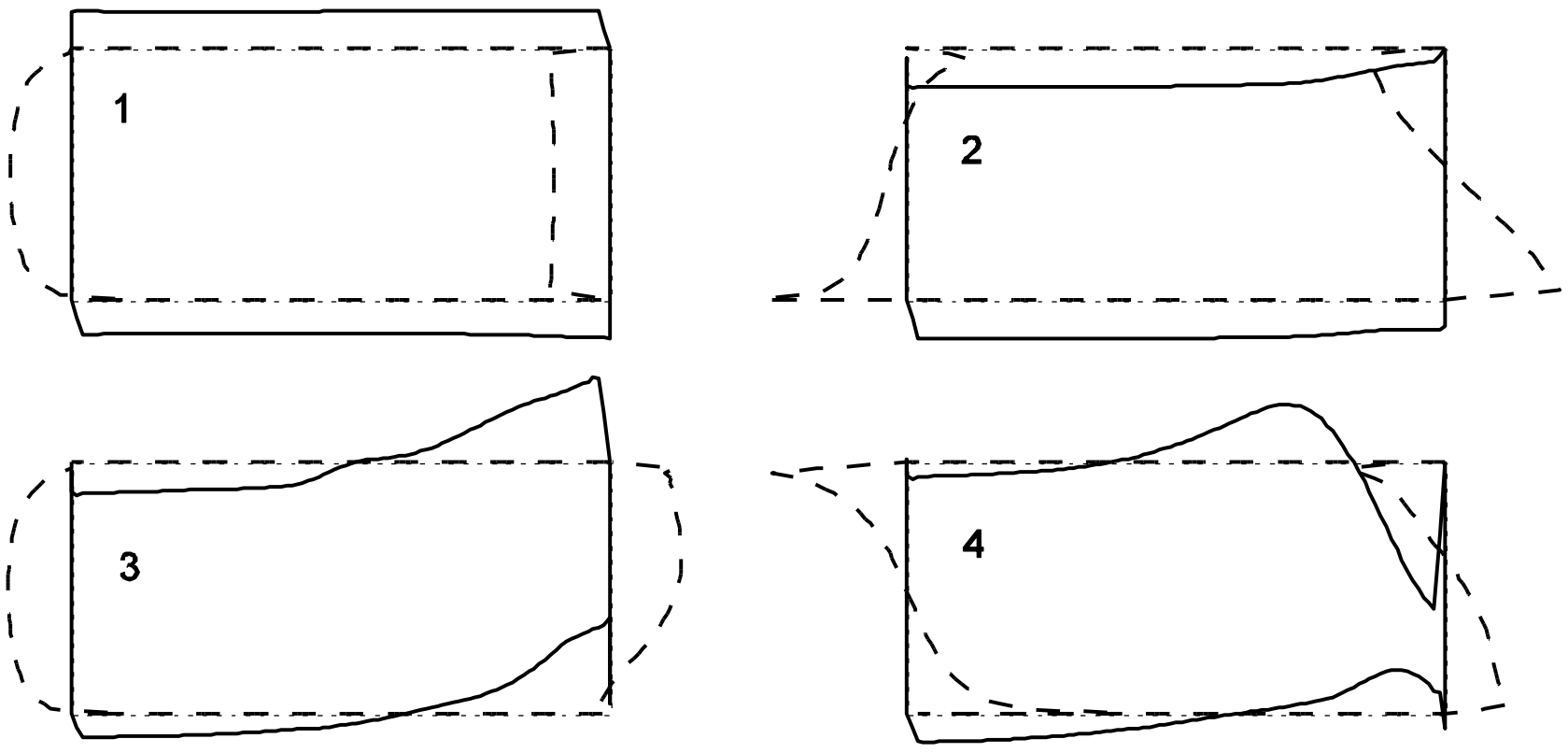

Figure 8.

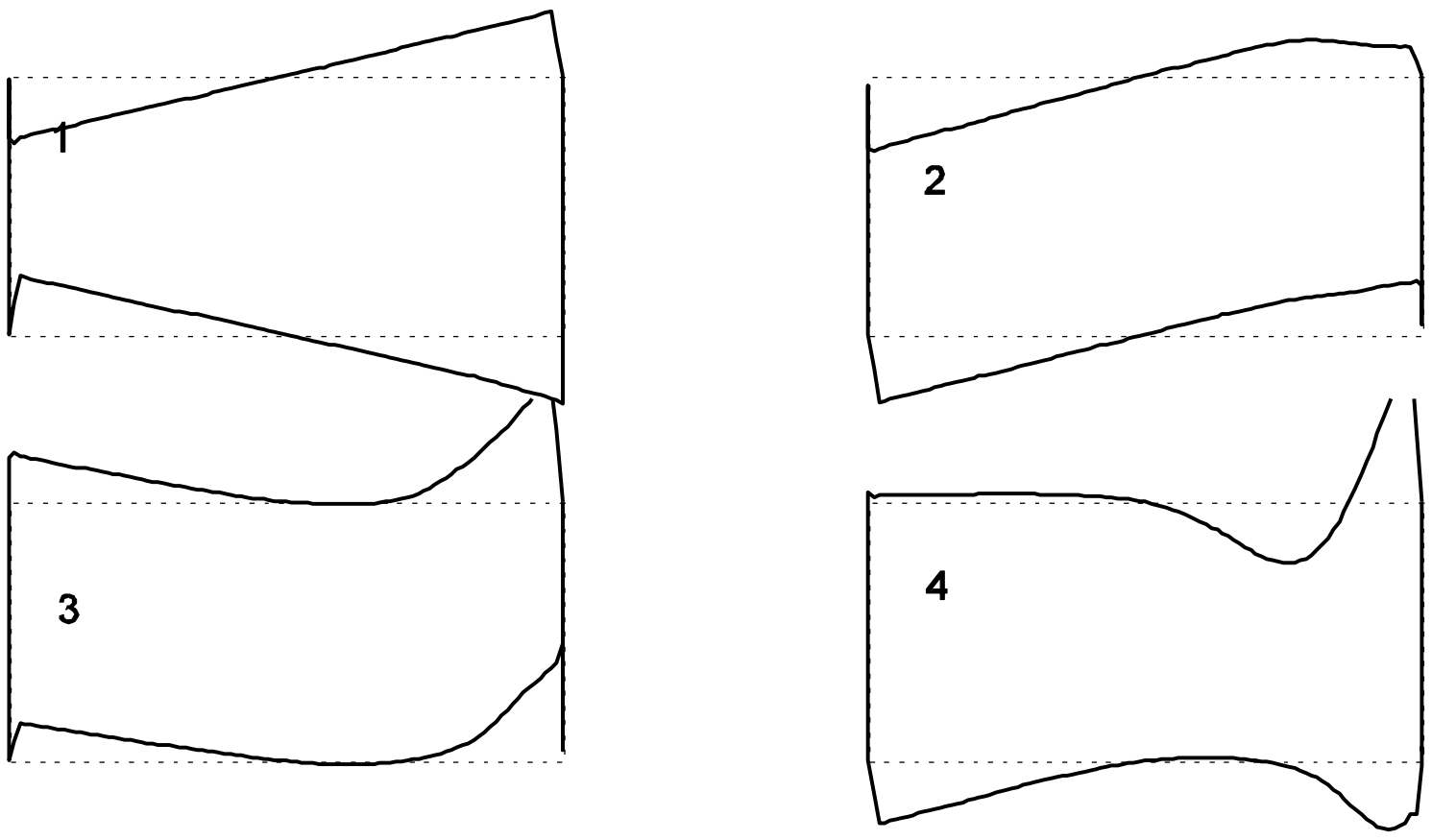

Figure 9.

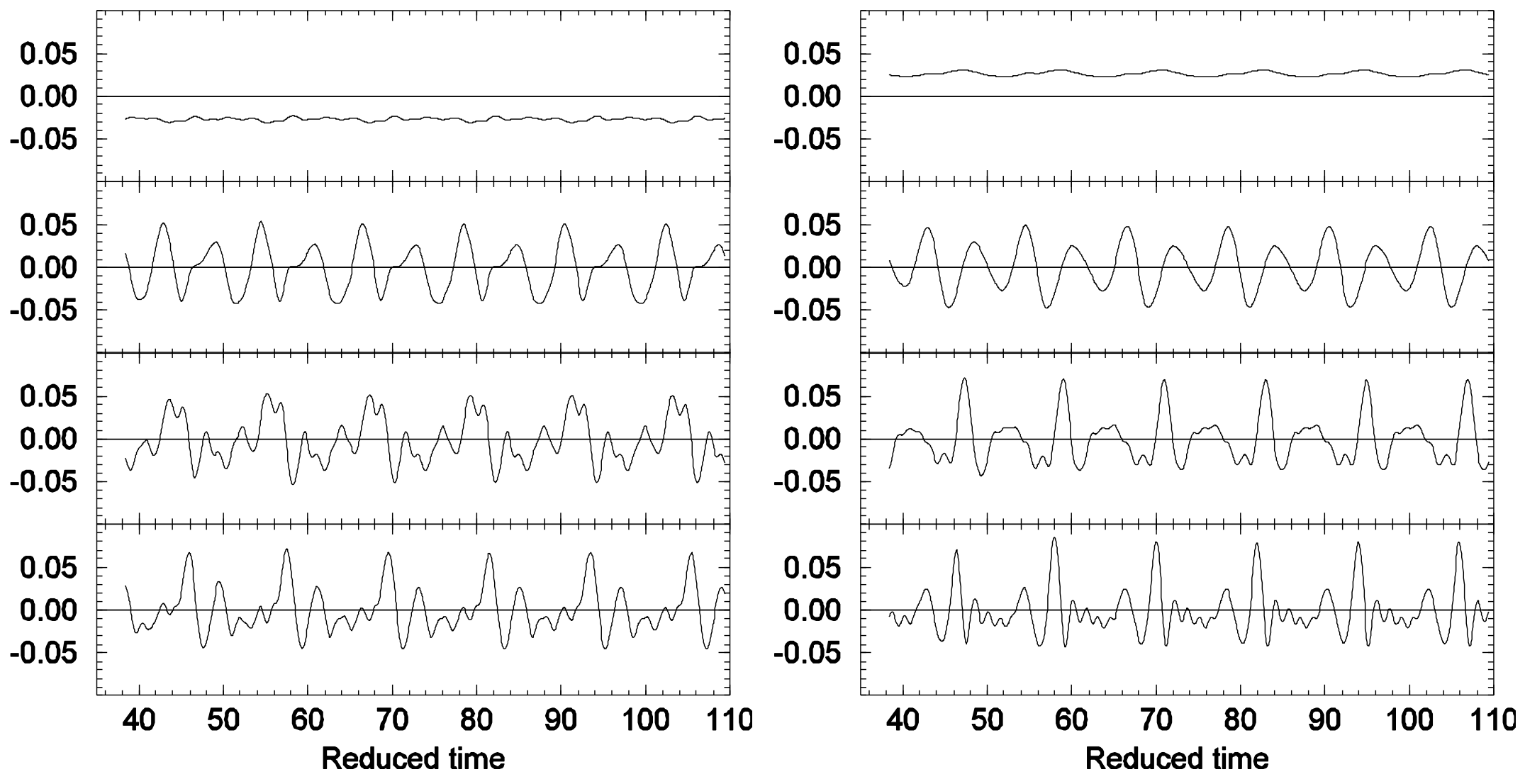

Figure 10.

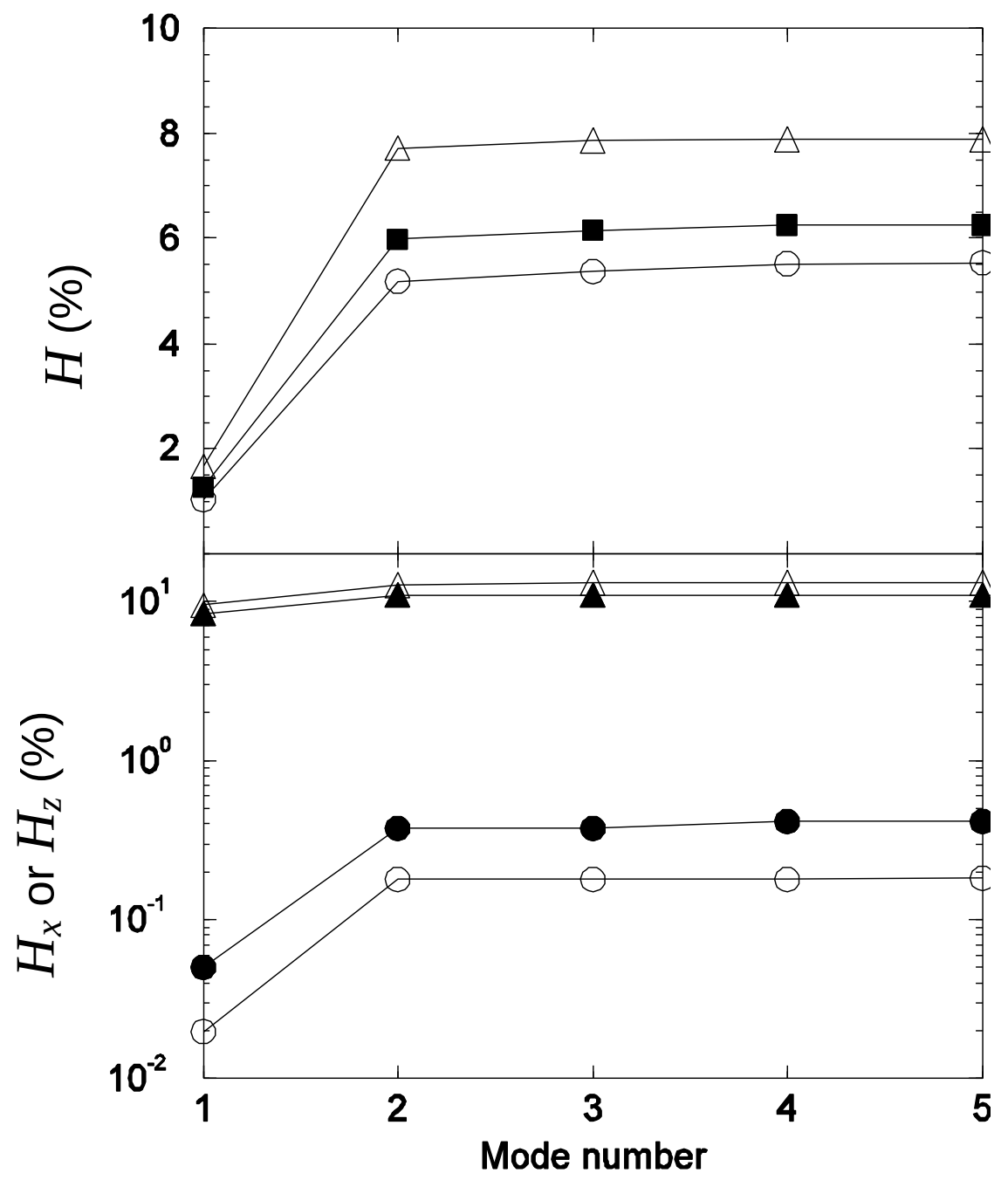

Figure 11.

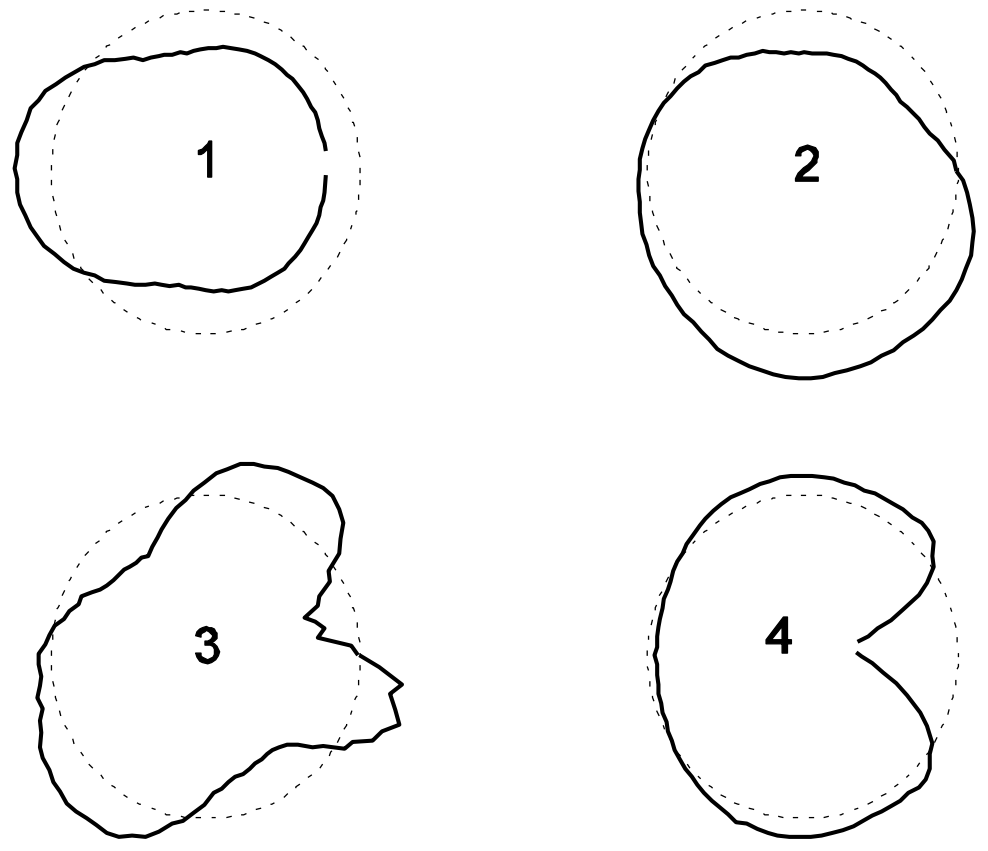

Figure 12.

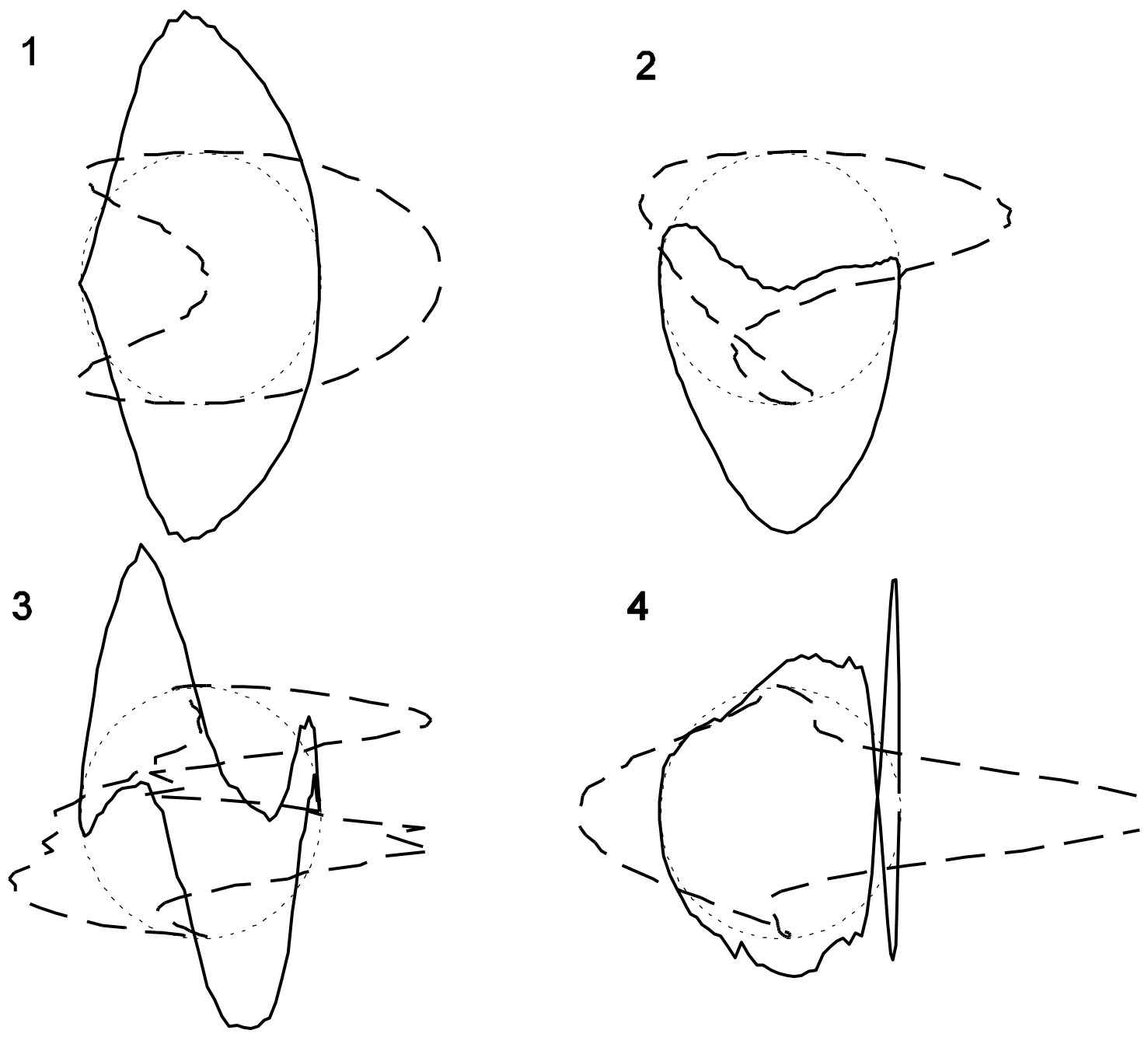

Figure 13.

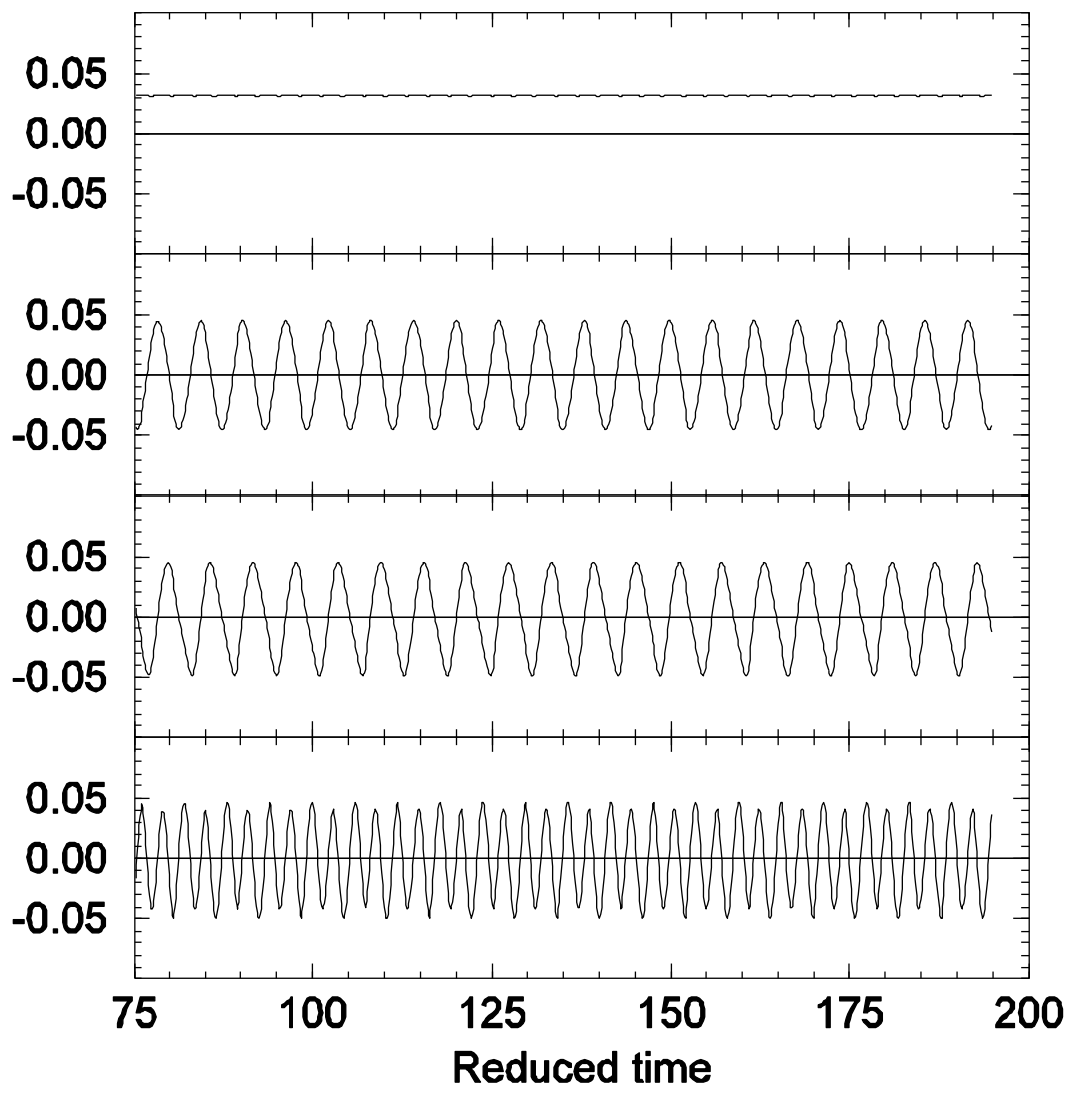

Figure 14.

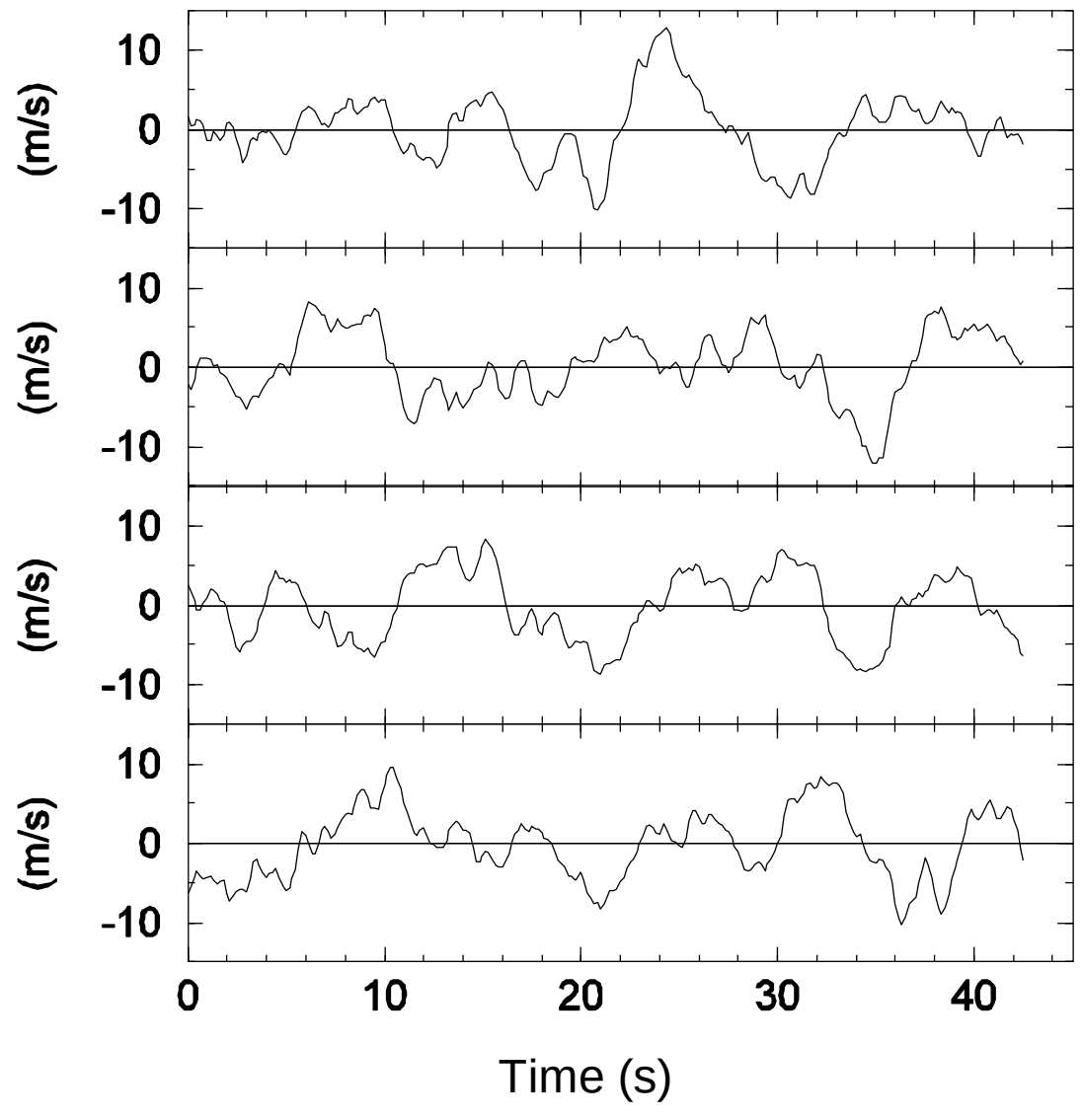

Figure 15.

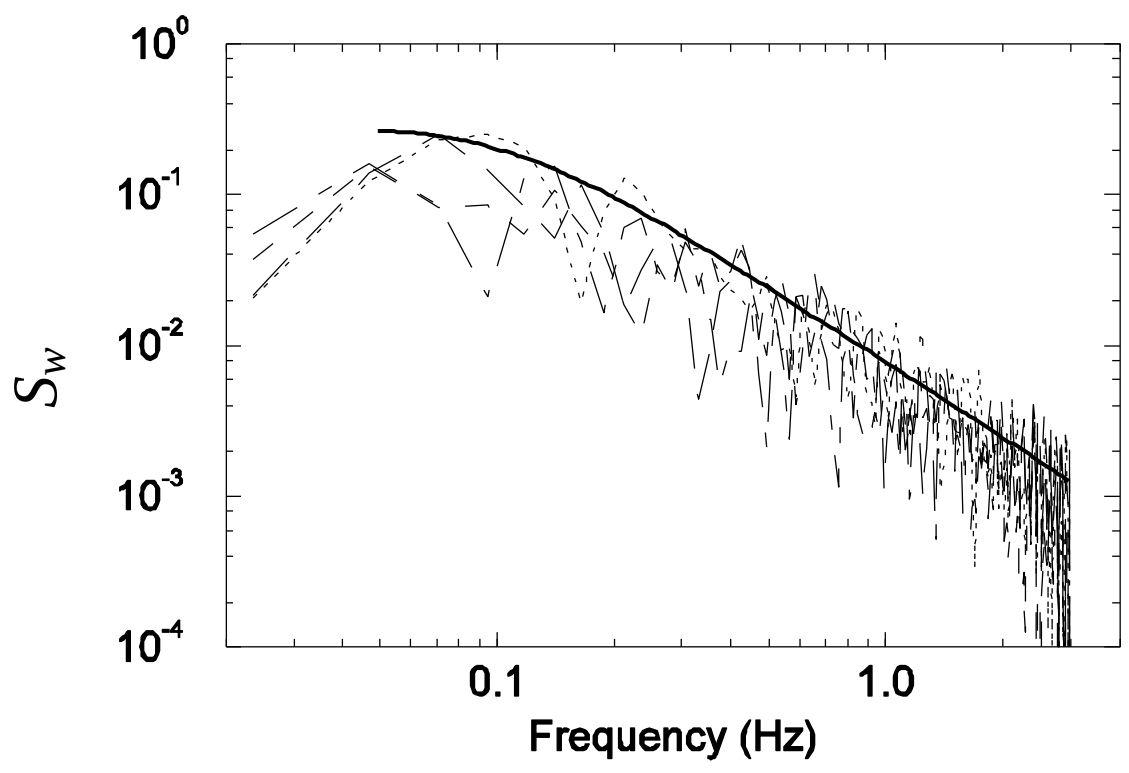

Figure 16.

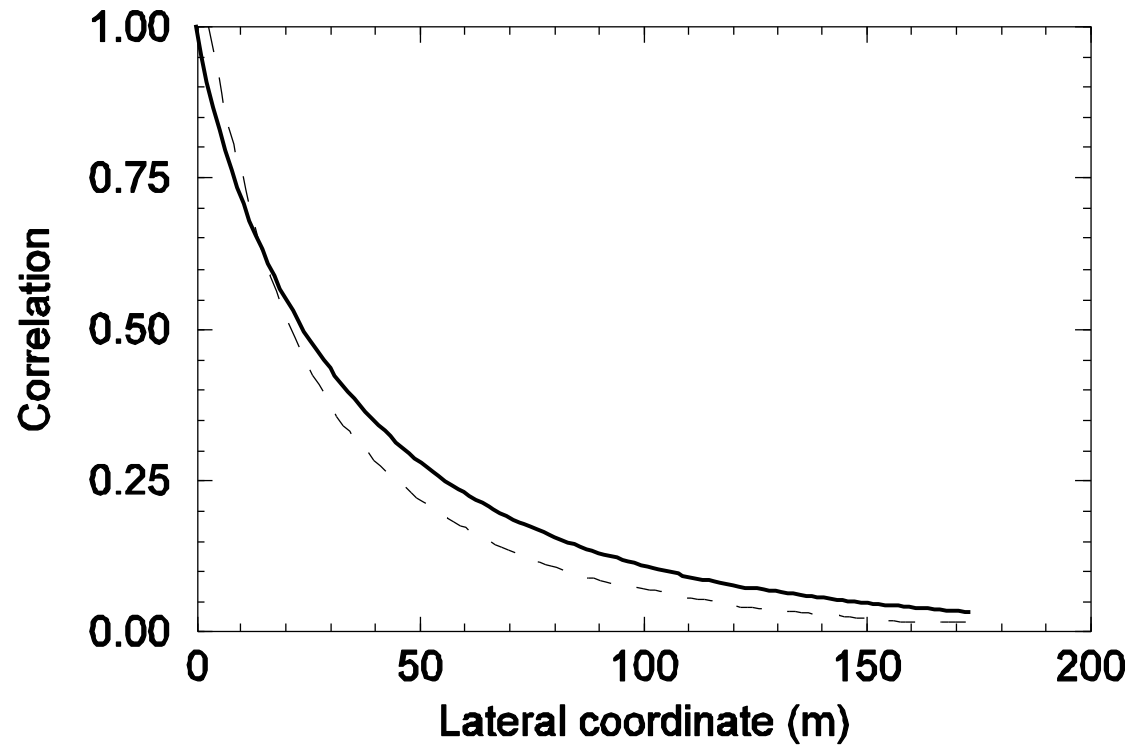

Figure 17.

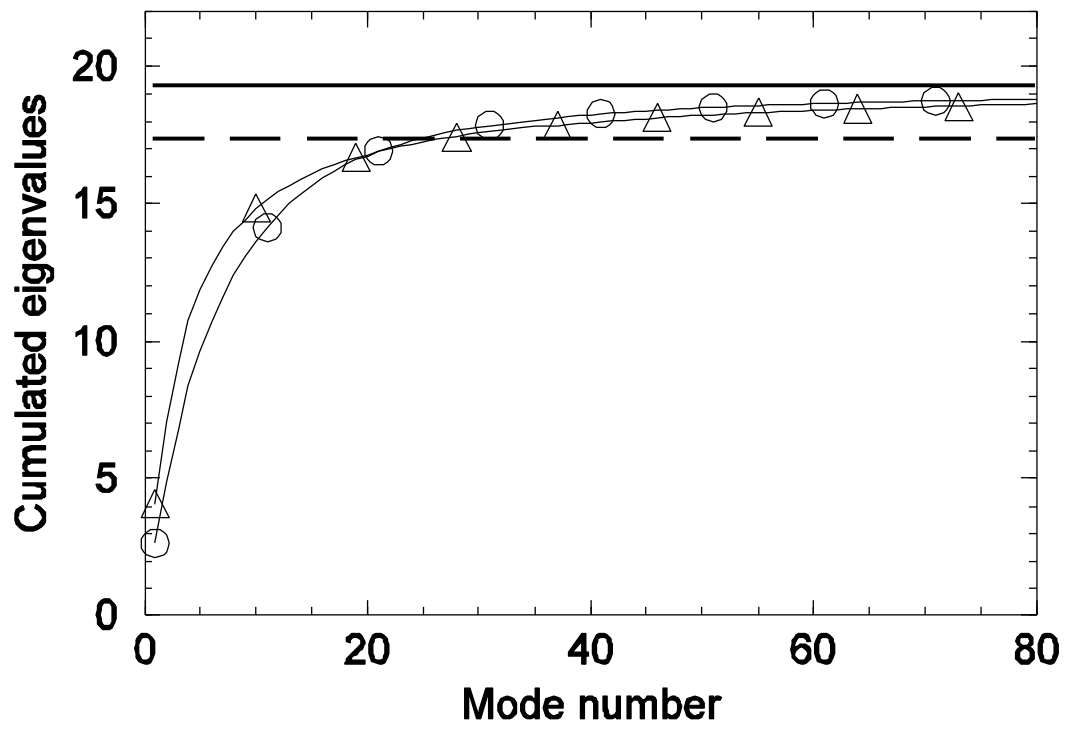

Figure 18.

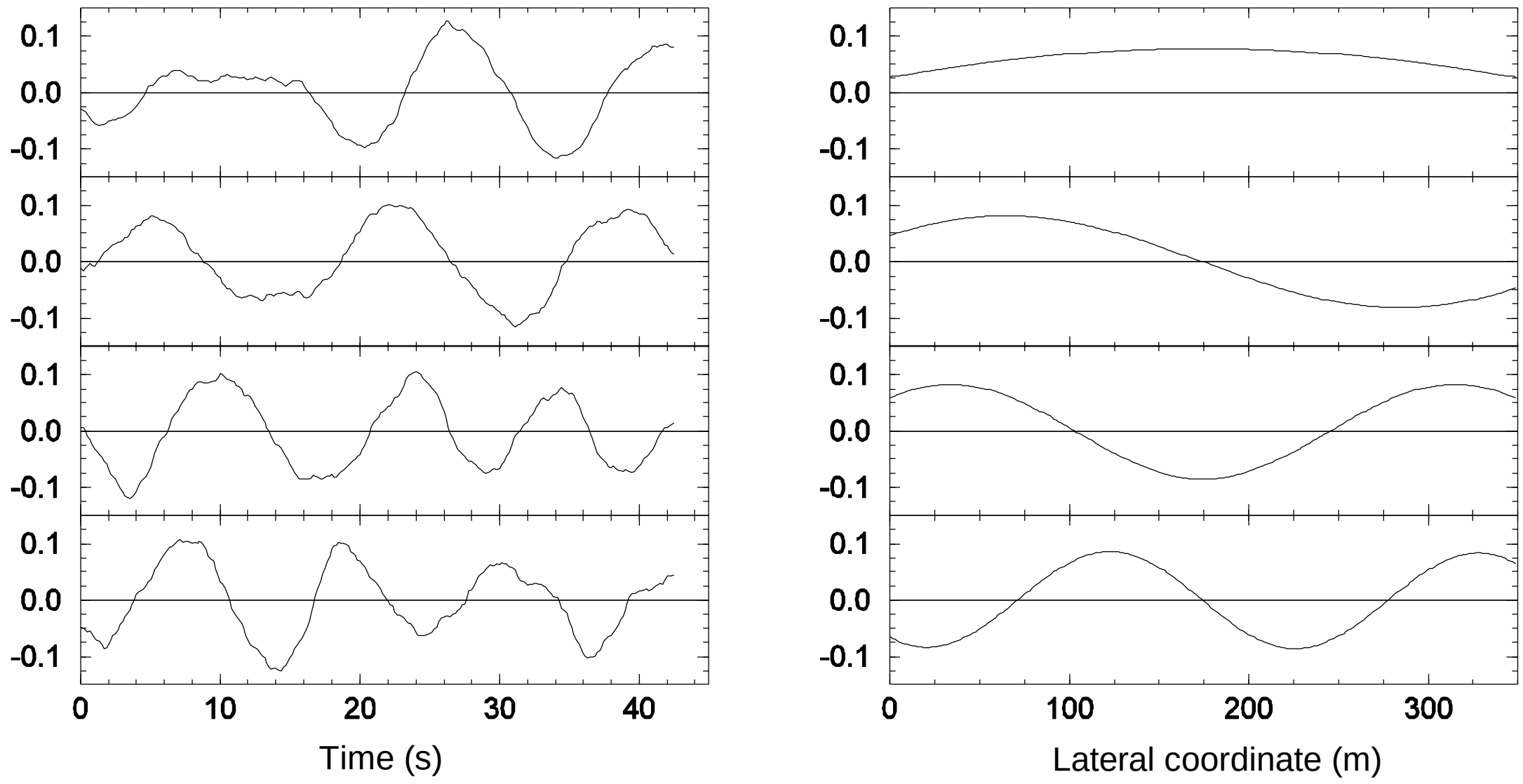

Figure 19.

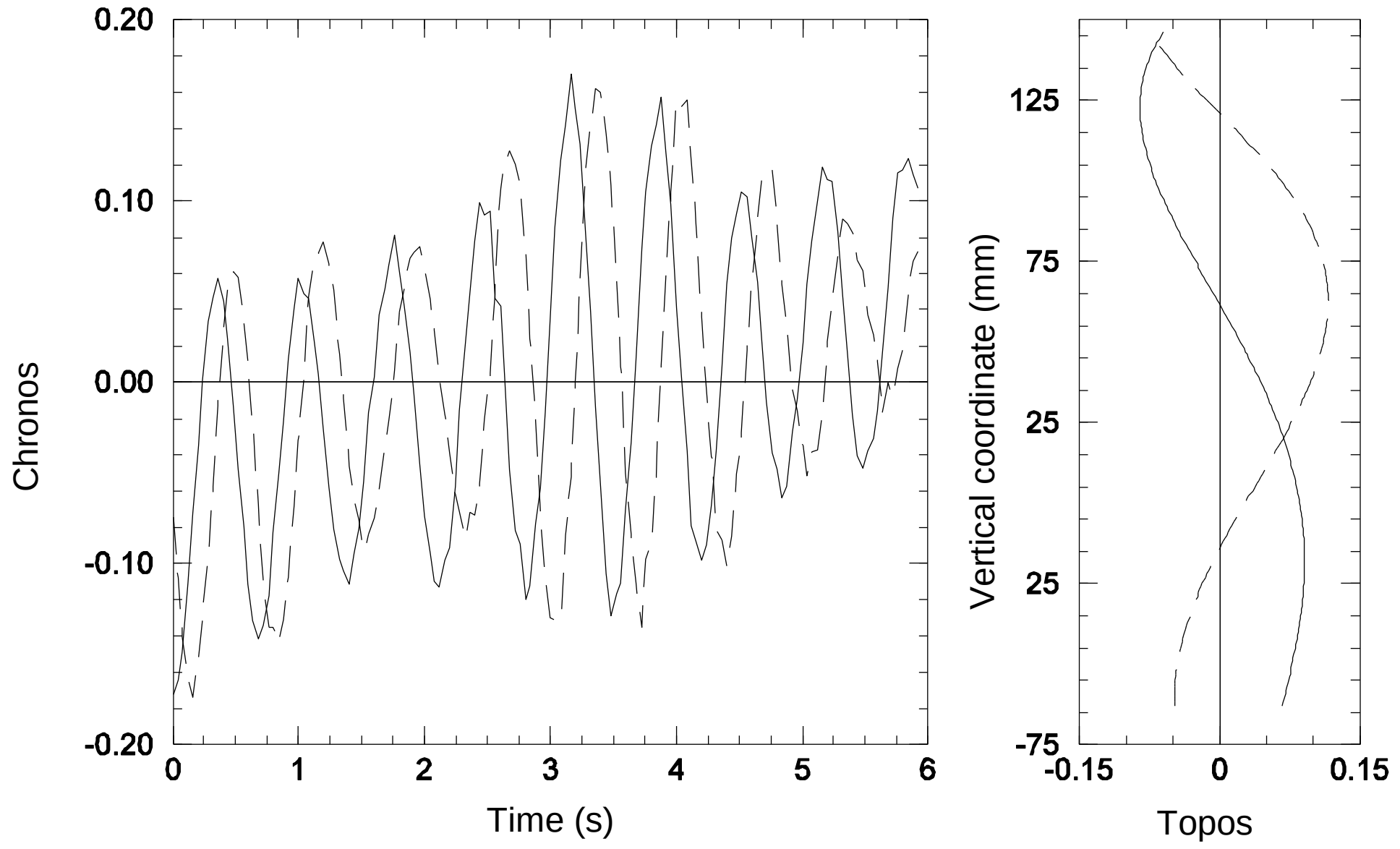

Figure 20.